\documentclass{acmtog} 
\acmVolume{VV}
\acmNumber{N}
\acmYear{YYYY}
\acmMonth{Month}
\acmArticleNum{XXX}
\doi{10.1145/XXXXXXX.YYYYYYY}
\setcopyright{rightsretained}
\def\BibTeX{{\rm B\kern-.05em{\sc i\kern-.025em b}\kern-.08em
    T\kern-.1667em\lower.7ex\hbox{E}\kern-.125emX}}

\title{Efficient Optimal Control of Smoke using Spacetime Multigrid}
\author{Zherong Pan \hspace*{8mm} Dinesh Manocha
        \thanks{\{zherong,dm\}@cs.unc.edu}\\ 
        Department of Computer Science\\ the University of North Carolina}

\usepackage{amsmath,amsfonts,amssymb}
\usepackage{graphicx}
\usepackage{prettyref}
\usepackage{mathrsfs}
\usepackage{wrapfig}
\usepackage{MnSymbol}
\usepackage{multirow}
\usepackage{xcolor}
\usepackage{booktabs}
\usepackage{algorithm,algpseudocode}
\usepackage{subfig}
\algnewcommand{\LineComment}[1]{\State \(\triangleright\) #1}

\newrefformat{Fig}{Figure~\ref{#1}}
\newrefformat{fig}{Figure~\ref{#1}}
\newrefformat{par}{Section~\ref{#1}}
\newrefformat{appen}{Appendix~\ref{#1}}
\newrefformat{sec}{Section~\ref{#1}}
\newrefformat{sub}{Section~\ref{#1}}
\newrefformat{table}{Table~\ref{#1}}
\newrefformat{alg}{Algorithm~\ref{#1}}
\newrefformat{Alg}{Algorithm~\ref{#1}}
\newrefformat{Def}{Definition~\ref{#1}}
\newrefformat{Thm}{Theorem~\ref{#1}}
\newrefformat{step}{Step~\ref{#1}}
\newrefformat{ln}{Line~\ref{#1}}
\newrefformat{eq}{Equation~\ref{#1}}
\newrefformat{pb}{Problem~\ref{#1}}
\newrefformat{it}{Item~\ref{#1}}
\newrefformat{te}{Term~\ref{#1}}
\def\Eqref Eq:#1:{\eqref{eq:#1}}
\newrefformat{Eq}{Equation~\Eqref#1:}

\newcommand{\E}[1]{\mathbf{#1}}
\newcommand{\TE}[1]{\textbf{#1}}

\newcommand{\FPP}[2]{\frac{\partial{#1}}{\partial{#2}}}

\newcommand{\THREER}[3]{\left(\setlength{\arraycolsep}{1pt}\begin{array}{ccc}{#1}^T & {#2}^T & {#3}^T\end{array}\right)^T}

\newcommand{\FOURC}[4]{\left(\begin{array}{c}#1 \\ #2 \\ #3 \\ #4\end{array}\right)}

\newcommand{\argmin}[1]{\underset{#1}{\E{argmin}}}

\newcommand{\ASelf}[1]{\E{Adv}\left[#1\right]}
\newcommand{\ARho}[2]{\E{A}\left[#1,#2\right]}

\usepackage{tikz}
\usetikzlibrary[arrows,backgrounds,decorations.pathmorphing,positioning,fit,petri]

\newcommand{\COMM}[2]{}{}
\newcommand{\FINE}[1]{}

\newcommand{\PO}{ Advection Optimization }
\newcommand{\AO}{ Navier-Stokes Optimization }
\newcommand{\AbbrPO}{ AO }
\newcommand{\AbbrAO}{ NSO }
\newcommand{\changed}[1]{\textcolor{black}{#1}}
\newenvironment{changedBlk}{\color{black}}{}

\begin{document}
\maketitle

\begin{abstract}
We present a novel algorithm to control the physically-based animation of smoke. Given a set of keyframe smoke shapes, we compute a dense sequence of control force fields that can drive the smoke shape to match several keyframes at certain time instances. Our approach formulates this control problem as a PDE constrained spacetime optimization and computes locally optimal control forces as the stationary point of the Karush-Kuhn-Tucker conditions. In order to reduce the high complexity of multiple passes of fluid resimulation, we utilize the coherence between consecutive fluid simulation passes and update our solution using a novel spacetime full approximation scheme (STFAS). We demonstrate the benefits of our approach by computing accurate solutions on 2D and 3D benchmarks. In practice, we observe more than an order of magnitude improvement over prior methods.
\end{abstract}

\section{Introduction}
Physically-based fluid animations are widely used in computer graphics and related areas. Over the past few years, research in fluid simulation has advanced considerably and it is now possible to generate plausible animations for movies and special effects in a few hours on current desktop systems. In this paper, we mainly deal with the problem of the keyframe-based spacetime control of smoke, a special kind of fluid. Given a set of keyframe smoke shapes, our goal is to compute a dense sequence of control forces such that the smoke can be driven to match these keyframes at certain time instances. This problem is an example of directable animation and arises in different applications, including special effects~\cite{rasmussen2004directable} (to model a character made of liquid) or artistic animation \cite{Angelidis:2006:CFS:1218064.1218068} (to change the moving direction of the smoke plume). Some of these control techniques, such as~\cite{nielsen2011guide}, are used in the commercial fluid software.

In practice, the keyframe-based control of fluids is still regarded as a challenging problem. Unlike fluid simulation, which deals with the problem of advancing the current fluid state to the next one by time integrating the Navier-Stokes equations, a fluid controller needs to consider an entire sequence of fluid states that results in a high dimensional space of possible control forces. For example, to control a 3D smoke animation discretized on a uniform grid at resolution $64^3$ with $60$ timesteps, the dimension of the resulting space of control forces can be as high as $10^8$. The problem of computing the appropriate control force sequence in such a high dimensional space can be challenging for any continuous optimization algorithm. Furthermore, the iterative computation of control forces 
would need many iterations, each of which involves solving a 2D or 3D fluid simulation problem that can take hours on a desktop system.

\begin{figure}[t]
\begin{center}
\includegraphics[trim={0cm 0cm 0cm 0cm},clip,width=0.45\textwidth]{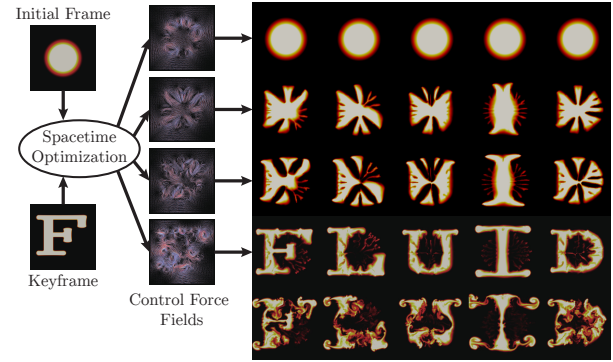}
\end{center}
\vspace{-5px}
\caption{\changed{\label{fig:Teaser} Given the keyframe, we use spacetime optimization to compute a dense sequence of control force fields, matching a smoke ball to the letter ``F'' (in the leftmost column). We highlight the control force fields. Five such animations are generated, matching the smoke ball to the word ``FLUID'', at resolution $128^2$ with $40$ timesteps. Each of these optimization computations take about half an hour on a desktop PC, and is about $17$ times faster than conventional gradient-based optimizer.}}
\vspace{-10px}
\end{figure}

Fluid control problems have been well studied in computer graphics and animation. At a broad level, prior techniques can be classified into proportional-derivative (PD) controllers and optimal controllers. PD controllers \cite{fattal2004target,shi2005taming} guide the fluid body using additional ghost force terms that are designed based on a distance measure between the current fluid shape and the keyframe. On the other hand, optimal controllers \cite{treuille2003keyframe,mcnamara2004fluid} formulate the problem as a spacetime optimization over the space of possible control forces constrained by the fluid governing equations, i.e., the Navier-Stokes equations. The objective function of this optimization formulation consists of two terms: The first term requires the fluid shape to match the keyframe shape at certain time instances, while the second term requires the control force magnitudes to be as small as possible. Optimal controllers are advantageous over PD controllers in that they search for the control forces with the smallest possible magnitude, which usually provide smoother keyframe transitions as well as satisfy the fluid dynamic constraints. Treuille et al.~\shortcite{treuille2003keyframe} and McNamara et al.~\shortcite{mcnamara2004fluid} use a simple gradient-based optimization to search for control forces. Although these techniques reduce the overhead by constraining the control forces to a small set of force templates, each gradient evaluation still needs to solve a fluid simulation problem, which can slow down the overall computation.

We present a new, efficient optimization algorithm for controlling smoke. Our approach exploits the special structure of the Navier-Stokes equations discretized on a regular staggered grid. The key idea is to solve the optimization problem by finding the stationary point of the first order optimality (Karush-Kuhn-Tucker) conditions \cite{nocedal2006numerical}. Unlike prior methods \cite{treuille2003keyframe,mcnamara2004fluid} that only solve for the primal variables, we maintain both the primal and dual variables (i.e., the Lagrangian multipliers). By maintaining the additional dual variables, we can iteratively update our solution without requiring it to satisfy the Navier-Stokes equations exactly in each iteration, thus avoiding repeated fluid simulation. In order to update the solution efficiently, we present a spacetime full approximation scheme (STFAS), which is a spacetime nonlinear multigrid solver. Our multigrid solver uses a novel spacetime smoothing operator and can converge within a number of iterations independent of the grid resolution and the number of timesteps. Overall, some novel aspects of our approach include:
\begin{itemize}
\item A spacetime optimization solver for controlling smoke animation. We formulate it using a fixed point iteration defined for the KKT conditions.
\item An acceleration scheme for the fixed point iteration using a spacetime full approximation scheme (STFAS).
\item A keyframe-based smoke control algorithm that can efficiently find high resolution control forces to direct long 2D and 3D fluid animation sequences. Moreover, a user can easily balance the keyframe matching exactness and the amount of smoke-like behavior by tuning a control force regularization parameter.
\end{itemize}

We have evaluated our approach on several benchmarks. Our benchmarks vary in terms of the grid resolution, the number of timesteps, and the control force regularization parameter. We highlight results with up to $60$ timesteps at the resolution of $64^3$. Moreover, we allow each component of the velocity field to be controlled. In practice, our algorithm can compute a convergent animation in less than $50$ iterations, and the overall runtime performance is about an order of magnitude faster than a gradient-based quasi-Newton optimizer \cite{nocedal2006numerical} for similar accuracy. An example of achieved smooth transitions between keyframes is illustrated in \prettyref{fig:Teaser}. 


\section{Related Work}\label{sec:related}
In this section, we give a brief overview of prior techniques for fluid simulation, multigrid solvers and animation control algorithms.

\TE{Fluid simulation} has been an active area of research in both computer graphics and computational fluid dynamics. The simulation of fluid is typically solved by a discretized time integration of the Navier-Stokes equations or their equivalent forms. At a broad level, prior fluid simulators can be classified into Lagrangian or Eulerian solvers according to the discretization of the convection operator. In order to model smoke and fire, a purely Eulerian solver \cite{fedkiw2001visual} is the standard technique. In terms of free-surface flow, hybrid Lagrangian-Eulerian representation \cite{zhu2005animating} has been widely used in computer graphics. In our work, we confine ourselves to the control of fluids without free-surface, i.e., smoke or fire. We use \cite{HarlowWelch1965} as our underlying fluid simulator.

\TE{Multigrid} solvers are widely used for fluid simulation. Multigrid is a long-standing concept that has been widely used to efficiently solve linear systems discretized from elliptic partial differential equations (see \cite{Trottenberg:2000:MUL:374106}). This idea has been successfully applied to fluid simulation~\cite{Chentanez:2007:LSL,chentanez2011real,zhang2014pppm} to find the solenoidal component of the velocity field. In terms of PDE-constrained optimization and control theory, the idea of multigrid acceleration has been extended to the spatial temporal domain. Borzi and Griesse~\shortcite{borzi2005experiences} proposed a semi-coarsening spacetime multigrid to control the time-dependent reaction-diffusion equation. Hinze et al.~\shortcite{hinze2012space} used a spacetime multigrid to solve the velocity tracking problem governed by the Navier-Stokes equations.

\TE{Fluid control} problems tend to be challenging and computationally demanding. Compared to other kinds of animations, e.g., character locomotion \cite{mordatch2012discovery}, the configuration space of fluid body is of much higher dimension. Prior work in this area can be classified into two categories: PD controllers \cite{fattal2004target,shi2005taming} and optimal controllers \cite{treuille2003keyframe,mcnamara2004fluid}. PD controllers compute the control forces by considering only the configuration of the fluid at the current and next time instance. For example, in \cite{shi2005taming}, a PD controller is used where the control forces are made proportional to the error between the current fluid shape and the target keyframe shape. Similar ideas are used for controlling smoke \cite{fattal2004target} and liquid \cite{shi2005taming,raveendran2012controlling}. In contrast, optimal controllers search for a sequence of control forces that minimize an objective function. Prior methods \cite{treuille2003keyframe,mcnamara2004fluid} typically solve spacetime optimization over a high-DOF search space to compute such control forces. Recently, these two methods have been combined~\cite{pan2013interactive} by first optimizing for the fluid shape at each keyframe and then propagating the changes to the neighboring timesteps. Fluid control can also be achieved by combining or modifying the results of existing fluid simulation data~\cite{raveendran2014blending} or guiding fluid using a designed low-resolution animation~\cite{nielsen2011guide,nielsen2010improved}. Our approach is also based on the spacetime optimization formulation, similar to \cite{treuille2003keyframe,mcnamara2004fluid}. 

\changed{In addition, there is considerable work on \TE{Fluid Capture}, which tries to digitize a fully or partially observed fluid animation. A row of methods have been developed for capturing fluid with specific appearance such as gas \cite{atcheson2008time} and flames \cite{ihrke2004image}, or capturing general flows \cite{Gregson:2014:CSC:2601097.2601147}. In many ways, the complexity of fluid capture problems lies between fluid simulation and fluid control problems. Although fluid capture problems can also be formulated as a spacetime optimization, most of the resulting algorithms do not take into account fluid dynamics as part of the formulation. Instead, they either assume that fluid dynamics have been captured in the observation, or reduce the tracking computation to a pure advection problem \cite{corpetti2000adaptation,kadri2013divergence}.}

\section{Fluid Control}\label{sec:define}

\begin{figure*}[t]
\begin{center}
\includegraphics[trim={0cm 0cm 0cm 0cm},clip,width=0.95\textwidth]{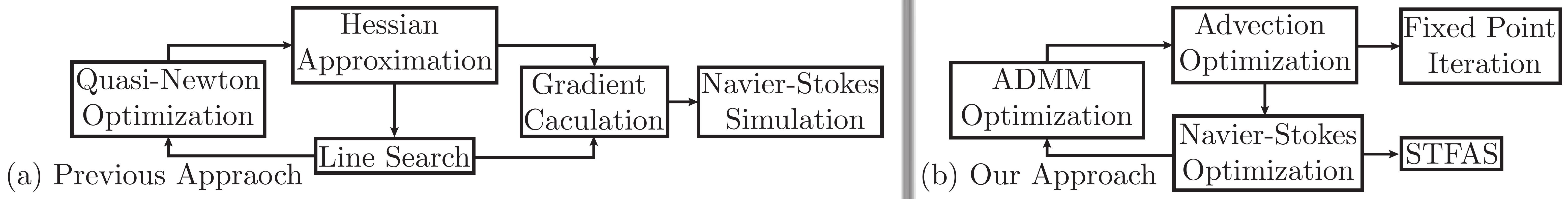}
\end{center}
\caption{\label{Fig:flowchart} A visual comparison of our algorithm pipeline and previous methods. Our key contribution is to avoid repeated and exact Navier-Stokes simulations. In our method, these simulations are replaced with a Navier-Stokes optimization, which is solved inexactly with a spacetime full approximation scheme (STFAS).}
\vspace{-15px}
\end{figure*}

In this section, we formulate the spacetime fluid control problem based on fluid dynamics (\prettyref{sec:Fluid}) and optimal control theory (\prettyref{sec:OCT}). The set of symbols used throughout the paper can be found in \prettyref{Fig:param}, and the subscript $i$ is the timestep index.
\begin{figure}[h]

\subfloat[][Symbols for fluid dynamic system]{\scalebox{0.65}{
\begin{tabular}{ll}
\hline  \\
Symbol & Meaning   \\
\hline  \\
$v_i$ & velocity field  \\
$u_i$ & ghost force field   \\
$\rho_i$ & density or dye field \\
$p_i$ & pressure field    \\
$s_i$ & state vector    \\
$\E{Adv}$ & self advection operator    \\
$\E{A}$ & passive advection operator  \\
$\Delta t$ & timestep size  \\
$N$ & number of timesteps
\vspace{39px}
\end{tabular}
}}
\hspace*{\fill}
\subfloat[][Symbols for spacetime optimization]{\scalebox{0.65}{
\begin{tabular}{ll}
\hline  \\
Symbol & Meaning   \\
\hline  \\
$v_i^*$ & slack velocity field  \\
$\lambda_i$ & augmented Lagrangian multiplier field   \\
$\mu_i$ & Lagrangian multiplier for passive advection  \\
$\gamma_i$ & Lagrangian multiplier for $\nabla\cdot v_i^*=0$  \\
$\bar{p}_i$ & Lagrangian multiplier for $\nabla\cdot v_i=0$ \\
$K$ & penalty coefficient for constraint $v_i=v_i^*$  \\
$r$ & regularization coefficient for $u_i$  \\
$c_i$ & indicator of keyframe at timestep $i$ \\
$C_i$ & metric measure for density field \\
$\E{R}$ & restriction operator of STFAS \\
$\E{P}$ & prolongation operator of STFAS    \\
$\E{S}$ & smoothing operator of STFAS   \\
$\E{Q}$ & solenoidal projection operator
\end{tabular}
}}
\vspace{-5px}
\caption{\label{Fig:param} Symbol table.}
\vspace{-5px}
\end{figure}
In general, we are dealing with a dynamic system whose configuration space is denoted as $s_i$ at physical time $i\Delta t$. Consecutive configurations $s_i$ and $s_{i+1}$ are related by the partial differential equation denoted as the function $f$: $s_{i+1}=f(s_i,u_i,\Delta t)$, where $u_i$ is the control input. An optimal controller computes a set of control inputs $\{u_i|i=0,\cdots,N-1\}$ that minimize the objective function denoted as function $E(s_0,\cdots,s_N)$. The overall optimal control problem is specified using the pair of functions $f$ and $E$. In the case of smoke control problems, $f$ is a discretization of the Navier-Stokes equations, and $E$ measures the difference between the smoke and keyframe shapes at certain time instances.

\subsection{Fluid Dynamic System\label{sec:Fluid}}
In our work, we restrict ourselves to the control of incompressible fluids without a free surface. Fluids such as smoke and fire, which are commonly used in movies and animations, fall into this category. We omit viscous terms for brevity. Small viscosity can be handled by a slight modification to $f$, which does not increase the complexity of our algorithm. Following \cite{HarlowWelch1965,pavlov2011structure}, we discretize the velocity-vorticity version of the Navier-Stokes equations using finite difference scheme and backward Euler time integrator for advection. Our configuration space $s_i=\THREER{v_i}{p_i}{\rho_i}$ concatenates the velocity field $v_i$, the kinetic pressure field $p_i$, and the density or dye field $\rho_i$. These scalar and vector fields are discretized on a staggered grid, which has been widely used by previous works such as \cite{fedkiw2001visual}. The transfer function $f$ under such discretization can be represented as:
\begin{align}
\label{eq:NSA}
&&\frac{v_{i+1}-v_i}{\Delta t}+\ASelf{v_{i+1}}=u_i-\nabla p_{i+1},  \\
\label{eq:NSB}
&&\nabla\cdot v_{i+1}=0, \\
\label{eq:Conv}
&&\rho_{i+1}=\ARho{\rho_i}{v_i},
\end{align}
where the self-advection operator $\ASelf{\bullet}$ is a discretization of the quadratic operator $\nabla\times\bullet\times\bullet$ and we assume constant unit fluid density. The pressure field $p_{i+1}$ is identified with the Lagrangian multiplier of the divergence free constraints $\nabla\cdot v_{i+1}=0$. Finally, the operator $\ARho{\bullet}{\bullet}$ is the passive scalar advection operator discretized as: $\rho_{i+1}=\E{e}^{\E{A}(v_i)\Delta t}\rho_{i}$, where matrix $\E{A}(v_i)$ is the second order upwinding stencil \cite{leonard1979stable}. By approximating the matrix exponential using Taylor series, the advection operator can be defined as:
\begin{align}
\label{eq:Adv}
\ARho{\rho_i}{v_i}=\sum_{k=0}^\infty\frac{\Delta t^k}{k!}\E{A}(v_i)^k\rho_i.
\end{align}
When $k$ tends to infinity, this upwinding advection operator is unconditionally stable since $\E{A}(v_i)$ is skew-symmetric, so that $\E{e}^{\E{A}(v_i)\Delta t}$ is an orthogonal matrix and $\|\rho_{i+1}\|=\|\rho_i\|$. In practice, we truncate $k$ to a finite value. Specifically, we set $k$ adaptively to be the smallest integer satisfying $\frac{\Delta t^k}{k!}\E{A}(v_i)^k\rho_i < 1e^{-5}$. \changed{Although this operator is computationally more expensive than the widely used semi-Lagrangian operator, it generates smoother controlled animations with large timestep size, as shown in \prettyref{fig:Popping}. This is useful when fewer timesteps are used to reduce the runtime cost.}

\subsection{Spacetime Optimization\label{sec:OCT}}
The optimal control of the dynamic system, discussed in \prettyref{sec:Fluid}, can be formulated as a spacetime optimization over the configuration trajectory $s_0,\cdots,s_N$. Our objective function is similar to the ones proposed in prior works~\cite{treuille2003keyframe,mcnamara2004fluid} that try to match $\rho_i$ to a set of keyframes $\rho_i^*$ while minimizing the magnitude of control forces $u_i$. The overall optimization problem can be formulated as:
\begin{align}
\label{eq:opt}
\argmin{u_i}&& \frac{1}{2}\sum_{i=0}^N c_i\|\rho_i-\rho_i^*\|^2+\frac{r}{2}\sum_{i=0}^{N-1}\|u_i\|^2   \\
\E{s.t.}&& s_{i+1}=f(s_i,u_i,\Delta t)\nonumber,
\end{align}
where $c_i$ is $1$ if there is a keyframe $\rho_i^*$ at frame $i$ and $0$ otherwise. $r$ is the regularization coefficient of the control forces. 

Treuille et al. \shortcite{treuille2003keyframe} and McNamara et al. \shortcite{mcnamara2004fluid} solve this optimization by eliminating the transfer function $f$ and plugging them into the objective function. Although this reformulation simplifies the problem into an unconstrained optimization, their new objective function takes a much more complex form, which is a long chain of function compositions. To minimize the new objective function, Treuille et al.~\shortcite{treuille2003keyframe} and McNamara et al.~\shortcite{mcnamara2004fluid} use a general-purpose gradient-based optimizer. As illustrated in \prettyref{Fig:flowchart} (a), a typical gradient-based optimizer such as the Quasi-Newton method requires repeated gradient calculation to approximate the Hessian matrix and performs line search to compute the stepsize. Each such gradient calculation requires a fluid resimulation, which becomes the major bottleneck in their algorithm.

\vspace{-4px}
\subsection{Our Approach}
\vspace{-4px}
Prior methods require that the solution computed during each iteration should satisfy the Navier-Stokes equations exactly, i.e., is a feasible solution. As a result, each iteration takes considerable running time. In practice, this requirement can be overly conservative because we only need to ensure that the final computed solution at the end of the algorithm is feasible. Thus, we can relax this requirement during the intermediate steps, and only need to ensure that the final solution lies in the feasible domain. This is a well-known idea and has been used by many other numerical optimization algorithms such as the interior point method \cite{nocedal2006numerical}.

Based on this observation, we design a new optimization pipeline illustrated in \prettyref{Fig:flowchart} (b). We first notice that our objective function is essentially constrained by two kinds of partial differential equations: the passive advection (\prettyref{eq:Conv}) governing the time evolution of the density field $\rho_i$; and the incompressible Navier-Stokes (\prettyref{eq:NSA} and \prettyref{eq:NSB}) governing the time evolution of the velocity field $v_i$. We introduce a set of slack variables to break these two kinds of constraints into two subproblems: \PO (\AbbrPO) is constrained only by \prettyref{eq:Conv} and \AO (\AbbrAO) is constrained only by \prettyref{eq:NSA} and \prettyref{eq:NSB}. In order to solve the \PO (\prettyref{sec:PO}), we use a fixed point iteration defined for its KKT conditions. For the \AO (\prettyref{sec:AO}), we update our solution using a spacetime full approximation scheme (STFAS) to avoid repeated fluid resimulations. This can lead to great speedup not only because of the fast convergence of a multigrid solver, but also because the multigrid solver allows warm-starting, so that we can make use of the coherence between consecutive iterations. In contrast, previous methods use fluid resimulations, which always solve Navier-Stokes equations from scratch, and solve them exactly.

\section{Spacetime Optimization using STFAS} \label{sec:tech}
In this section, we present our novel algorithm to solve \prettyref{eq:opt}. We also describe our new acceleration method, STFAS. \changed{See \cite{nocedal2006numerical} for an introduction to some notations and reference solvers used in this section.}

By introducing a series of slack variables $v_i^*$, we can decompose the overall optimization problem into two subproblems and reformulate \prettyref{eq:opt} as:
\begin{align}
\label{eq:optAdmm}
\argmin{u_i}&& \frac{1}{2}\sum_{i=0}^Nc_i\|\rho_i-\rho_i^*\|^2+\frac{r}{2}\sum_{i=0}^{N-1}\|u_i\|^2+\nonumber   \\
&&\lambda_i^T(v_i-v_i^*)+\frac{K}{2}\sum_{i=0}^{N-1}\|v_i-v_i^*\|^2  \\
\E{s.t.}&&\frac{v_{i+1}-v_i}{\Delta t}+\ASelf{v_{i+1}}=u_i-\nabla p_{i+1}\nonumber   \\
&&\rho_{i+1}=\ARho{\rho_i}{v_i^*},\quad \nabla\cdot v_i=0\nonumber,
\end{align}
where we added the augmented Lagrangian term $\lambda_i^T(v_i-v_i^*)$ and the penalty term $\frac{K}{2}\sum_{i=0}^{N-1}\|v_i-v_i^*\|^2$. This kind of optimization can be solved efficiently using the well-known alternating direction method of multipliers (ADMM) \cite{boyd2011distributed}, \changed{which has been used in \cite{Gregson:2014:CSC:2601097.2601147} for fluid tracking.} Specifically, in each iteration of our algorithm, we first fix $v_i,p_i$ and solve for $v_i^*$. This subproblem is denoted as the \PO (\AbbrPO) because the PDE constraints are just a passive advection of the density field $\rho_i$. We then fix $v_i^*$ and solve for $v_i,p_i$. We denote this subproblem as the \AO (\AbbrAO), constrained by the incompressible Navier-Stokes equations. The final step is to adjust $\lambda_i$ according to the constraint violation as: $\lambda_i=\lambda_i+K\beta(v_i-v_i^*)$ where $\beta$ is a constant parameter.

The advantage of breaking the problem up is that we can derive simple and effective algorithms to solve each subproblem. Our algorithm directly solves the first order optimality (KKT) conditions of both problems. To solve the \AbbrPO subproblem, we introduce a fixed point iteration in \prettyref{sec:PO}, while for the \AbbrAO subproblem, which is the bottleneck of the algorithm, we introduce STFAS solver in \prettyref{sec:AO}.

\subsection{\PO\label{sec:PO}}
The goal of solving the \AbbrPO subproblem is to find a sequence of velocity fields $v_i^*$ to advect $\rho_i$ so that it matches the keyframes, assuming that these $v_i^*$ are uncorrelated. By dropping terms irrelevant to $v_i^*$ from \prettyref{eq:optAdmm}, we get a concise formulation for the \AbbrPO subproblem:
\begin{align}
\label{eq:PO}
\argmin{v_i^*}&& \frac{1}{2}\sum_{i=0}^N\|\rho_i-\rho_i^*\|_{C_i}^2+\frac{K}{2}\sum_{i=0}^{N-1}\|v_i-v_i^*\|^2  \\
\E{s.t.}&&\rho_{i+1}=\ARho{\rho_i}{v_i^*}\quad \nabla\cdot v_i^*=0   \nonumber,
\end{align}
where we have absorbed the augmented Lagrangian term $\lambda_i^T(v_i-v_i^*)$ by setting: $v_i=v_i+\lambda_i/K$. 

Due to the inherent nonlinearity and ambiguity in the advection operator, an \AbbrPO solver is prone to falling into local minimum, leading to trivial solutions. We introduce two additional modifications to \prettyref{eq:PO} to avoid these trivial solutions. First, we replace the scalar coefficient $c_i$ with a matrix $C_i$ which could be used to avoid the problem of a zero gradient if the keyframe $\rho_i^*$ is far from the given density field $\rho_i$. \changed{Similar to \cite{treuille2003keyframe,fattal2004target}, we take $C_i=c_i\Sigma_iG_i^TG_i$ to be a series of Gaussian filters $G_i$ with receding support. Specifically, $G_i$ has a standard deviation $\sigma(G_i)=2\sigma(G_{i-1})$. The combination of these filters spreads the gradient information throughout the domain. Moreover, the keyframe features of various frequencies get equally penalized.}
We also introduce additional solenoidal constraints on $v_i^*$. Note that this term does not alter the optima of \prettyref{eq:optAdmm} since $v_i=v_i^*$ on convergence. However, it prevents the optimizer from creating or removing densities in order to match the keyframe, which is a tempting trivial solution.

We solve this optimization via a fixed point iteration derived from its KKT conditions. To derive this system we introduce Lagrangian multipliers $\mu_i$ for each advection equation $\rho_{i+1}=\ARho{\rho_i}{v_i^*}$ and $\gamma_i$ for the solenoidal constraints, giving a Lagrangian function:
\begin{align*}
&&\mathcal{L}=\frac{1}{2}\sum_{i=0}^N \|\rho_i-\rho_i^*\|_{C_i}^2+\frac{K}{2}\sum_{i=0}^{N-1} \|v_i-v_i^*\|^2+   \\
&&\sum_{i=0}^{N-1}\mu_i^T(\rho_{i+1}-\ARho{\rho_i}{v_i^*})+\gamma_i^T\nabla\cdot v_i^*.
\end{align*}
After taking the derivative of the above Lagrangian against $\rho_i,v_i^*$ (primal variables) and $\mu_i,\gamma_i$ (dual variables), respectively, we get the following set of KKT conditions for $0\leq i\leq N$:
\begin{align}
\label{eq:KKTPO}
&&\mu_{i-1}=\FPP{\ARho{\rho_i}{v_i^*}}{\rho_i}^T\mu_i-C_i(\rho_i-\rho_i^*) \nonumber  \\
&&v_{i-1}^*=\E{Q}(v_{i-1}+\FPP{\ARho{\rho_{i-1}}{v_{i-1}^*}}{v_{i-1}^*}^T\frac{\mu_{i-1}}{K})   \\
&&\rho_{i+1}-\ARho{\rho_i}{v_i^*}=0,\quad \nabla\cdot v_i^*=0 \nonumber,
\end{align}
where we set $\mu_{-1}=\mu_N=0$ to unify the index, and we have replaced $\gamma_i$ with a solenoidal projection operator $\E{Q}$. This actually defines a fixed point iteration where we can first update $\rho_i$ in a forward pass and then update $\mu_i, v_i$ in a backward pass. This is closely related to the adjoint method \cite{mcnamara2004fluid}, which also takes a forward-backward form. Unlike \cite{mcnamara2004fluid} which then solves $v_i^*$ using quasi-Newton method, a fixed point iteration is much simpler to implement, and a general-purpose optimizer is not needed. The mostly costly step in applying \prettyref{eq:KKTPO} is the operator $\E{Q}$ where we use conventional multigrid Poisson solver \cite{Trottenberg:2000:MUL:374106}. \changed{A pseudo-code of our \AbbrPO solver is given in \prettyref{Alg:Passive}.}
\setlength{\textfloatsep}{5pt}
\begin{algorithm}[h]
\caption{\label{Alg:Passive} The Fixed Point Iteration: This is used to solve the \AbbrPO subproblem. The algorithm consists of a forward sweep that updates the density fields $\rho_i$ and a backward sweep that updates $\mu_i$ and $v_i$.}
\begin{algorithmic}[1]
\Require Initial $v_i, \rho_0$ and keyframes $\rho_i^*$
\Ensure Fixed point $v_i, \mu_i$
\For{$i=0,\cdots,N-1$}
\LineComment{Initialization}
\State $v_i^*=v_i$
\EndFor
\For{$i=1,\cdots,N$}
\LineComment{Find primal variable $\rho$}
\State $\rho_i=\ARho{\rho_{i-1}}{v_{i-1}^*}$    
\EndFor
\State set $\mu_{-1}=\mu_N=0$
\For{$i=N,\cdots,1$}
\LineComment{Find dual variable $\mu$}
\State $\mu_{i-1}=\FPP{\ARho{\rho_i}{v_i^*})}{\rho_i}^T\mu_i-C_i(\rho_i-\rho_i^*)$  \label{ln:AdvInt}
\LineComment{Find primal variable $v$}
\State $v_{i-1}^*=\E{Q}(v_{i-1}+\FPP{\ARho{\rho_{i-1}}{v_{i-1}^*})}{v_{i-1}^*}^T\frac{\mu_{i-1}}{K})$ \label{ln:Proj}
\EndFor
\end{algorithmic}
\end{algorithm}

\begin{figure}
\begin{center}
\scalebox{0.9}{
\includegraphics[trim={0cm 0cm 0cm 0cm},clip,width=0.47\textwidth]{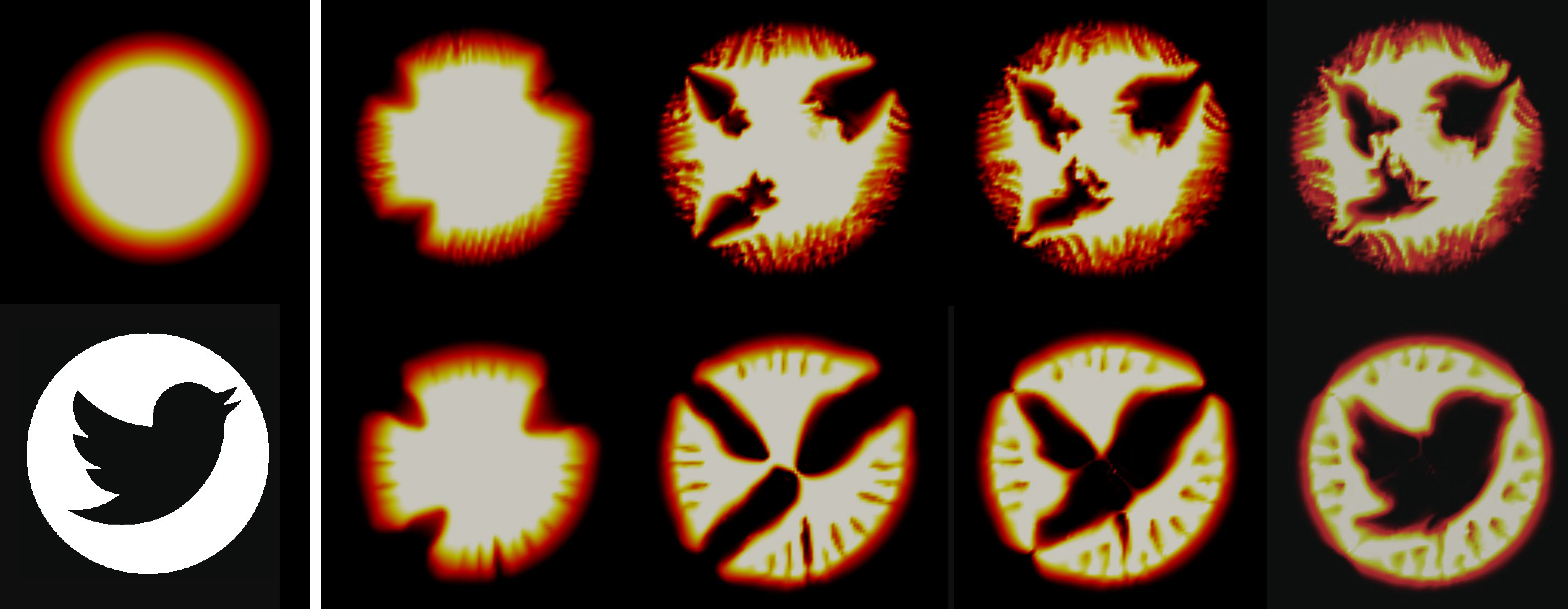}
\put(-175,-10){$t=3s$}
\put(-127,-10){$t=6s$}
\put(-80,-10){$t=9s$}
\put(-34,-10){$t=12s$}
}
\end{center}
\vspace{-5px}
\caption{\changed{\label{fig:Popping} We tested the fixed point iteration \prettyref{eq:KKTPO} using different advection operator $\ARho{\bullet}{\bullet}$ to deform an initially circle-shaped smoke (top left) into the bird icon (bottom left). The \AbbrPO subproblem solved using the semi-Lagrangian operator involves lots of popping artifacts (top row). The upwinding operator in \prettyref{eq:Adv} doesn't suffer from such problems (bottom row).}}
\end{figure}
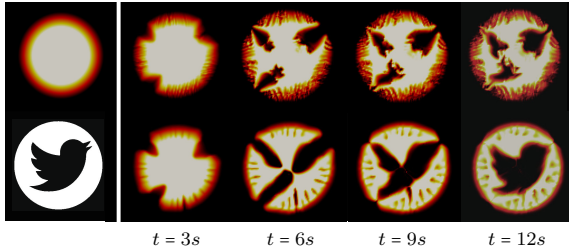

In the above derivation, since we do not exploit any structure in the operator $\ARho{\rho_i}{v_i^*}$, basically any advection operator other than \prettyref{eq:Adv}, such as semi-Lagrangian, could be used as long as its partial derivatives against $\rho_i,v_i^*$ are available. Empirically, however, \prettyref{eq:Adv} generally gives smoother animations especially under large timestep size. \changed{This is because the semi-Lagrangian operator can jump across multiple cells when performing backtracking, and the density value changes in these cells are ignored. As a result, the semi-Lagrangian operator suffers from popping artifacts as illustrated in \prettyref{fig:Popping}, while our operator (\prettyref{eq:Adv}), being purely grid-based, doesn't exhibit such problem. Unlike \cite{treuille2003keyframe}, where these popping artifacts can be alleviated by constraining control force fields to a small set of force templates, we allow every velocity component to be controlled. In this case, the use of our new advection operator is highly recommended.}

\subsection{\AO \label{sec:AO}}
Complementary to \prettyref{sec:PO}, the goal of the \AO is to enforce the correlation between $v_i$ given the sequence of guiding velocity fields $v_i^*$. The optimization takes the following form:
\begin{align*}
\argmin{v_i}&& \frac{r}{2}\sum_{i=0}^{N-1}\|u_i\|^2+\frac{K}{2}\sum_{i=0}^{N-1}\|v_i-v_i^*\|^2  \\
\E{s.t.}&&\frac{v_{i+1}-v_i}{\Delta t}+\ASelf{v_{i+1}}=u_i-\nabla p_{i+1} \\
&&\nabla\cdot v_i=0.
\end{align*}
This subproblem is the bottleneck of our algorithm, for which a forward-backward adjoint method similar to \prettyref{eq:KKTPO} requires solving the Navier-Stokes equations exactly in the forward pass. To avoid this costly solve, we update primal as well as dual variables in a single unified algorithm. In the same way as in \prettyref{sec:PO}, we derive the KKT conditions and assemble them into a set of nonlinear equations:
\begin{align}
\label{eq:KKTAO}
f=\FOURC
{\frac{K}{r}(v_i-v_i^*)+\FPP{u_i}{v_i}^Tu_i+\FPP{u_{i-1}}{v_i}^Tu_{i-1}+\nabla \bar{p}_i}
{\nabla\cdot v_i}
{\frac{v_{i+1}-v_i}{\Delta t}+\ASelf{v_{i+1}}-u_i+\nabla p_{i+1}}
{\nabla\cdot u_i}=0,
\end{align}
where the partial derivatives are $\FPP{u_i}{v_i}=-\frac{I}{\Delta t}$, $\FPP{u_{i-1}}{v_i}=\frac{I}{\Delta t}+\FPP{\ASelf{v_i}}{v_i}$, and the additional variable $\bar{p}_i$ is the Lagrangian multiplier for the solenoidal constraint: $\nabla\cdot v_i=0$. \changed{We refer readers to \prettyref{appen:appenA} for the derivation of \prettyref{eq:KKTAO}.} In summary, we have to solve for the primal variables $u_i,v_i$ as well as the dual variables $p_i,\bar{p}_i$. Unlike \prettyref{eq:KKTPO}, however, we do not differentiate these two sets of variables and solve for them by iteratively bringing the residual $f$ to zero.

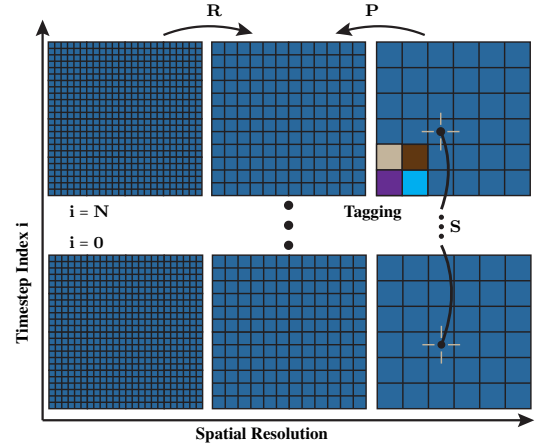
\begin{figure}[h]
\vspace{-5px}
\begin{center}
\scalebox{0.8}{
\includegraphics[width=0.47\textwidth]{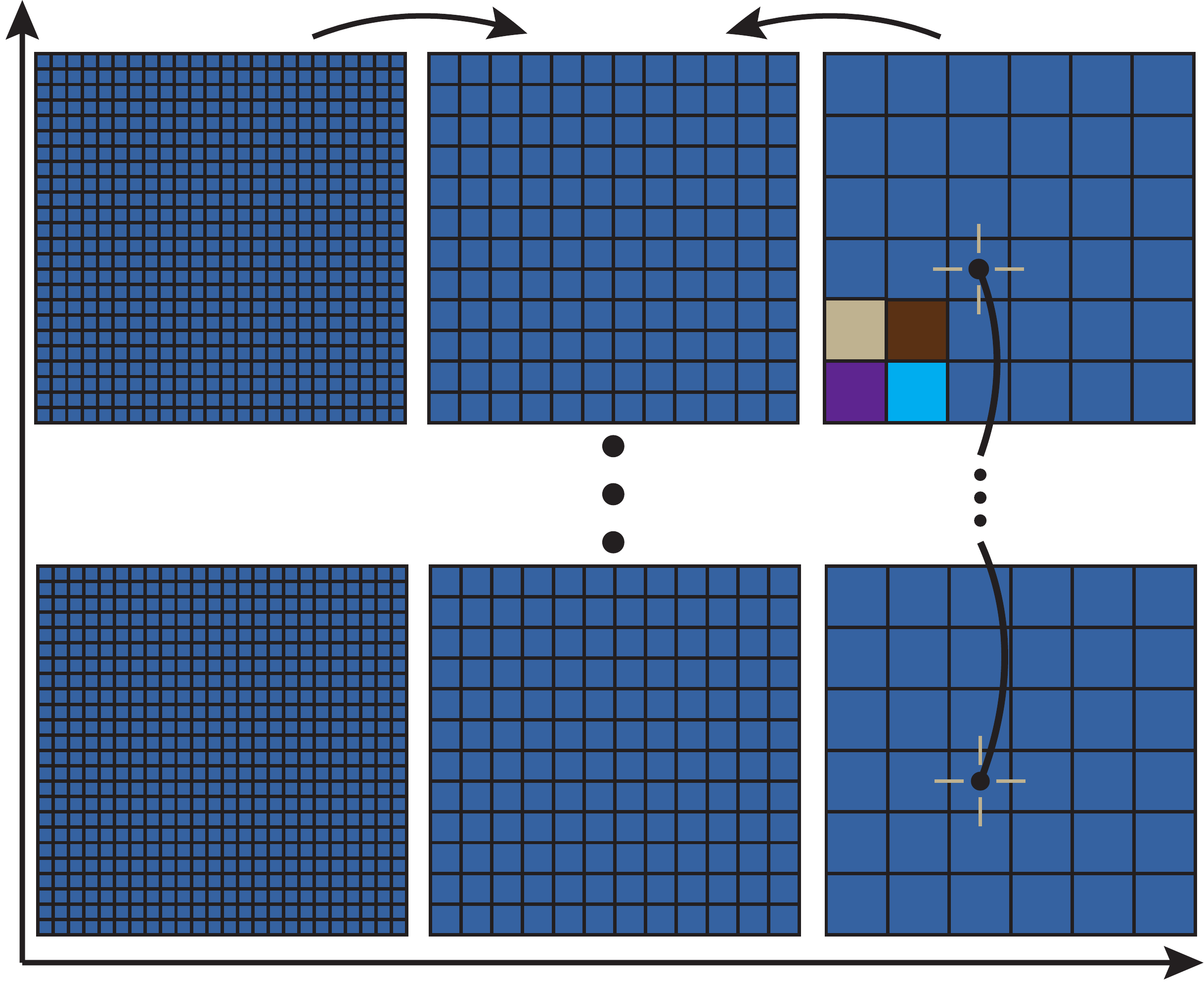}
\put(-220,100){$\E{i=N}$}
\put(-220,85){$\E{i=0}$}
\put(-155,195){$\E{R}$}
\put(-80,195){$\E{P}$}
\put(-40,93){$\E{S}$}
\put(-90,100){\TE{Tagging}}
\put(-160,-5){\TE{Spatial Resolution}}
\put(-245,40){\rotatebox{90}{\TE{Timestep Index i}}}
}
\end{center}
\caption{\label{fig:STFAS} A 2D illustration of our STFAS multigrid scheme. We use semi-coarsening only in the spatial direction (horizontal), with each finer level doubling the grid resolution. We use trilinear interpolation operators for $\E{P}, \E{R}$ and tridiagonal SCGS smoothing for $\E{S}$, which solves the primal variables $v_i,u_i$ (defined on faces as short white lines) and dual variables $p_i,\bar{p}_i$ (defined in cell centers as black dots) associated with one cell across all the timesteps (vertical) by solving a block tridiagonal system. The solve can be made parallel by the $8-color$ tagging in 3D or $4-color$ tagging in 2D.}
\vspace{-5px}
\end{figure}

To this end, we develop a spacetime full approximation scheme (STFAS), which is a geometric multigrid algorithm designed for solving a nonlinear system of equations as illustrated in \prettyref{fig:STFAS}. The multigrid solver is a classical tool originally used for solving linear systems induced from elliptical PDEs. We refer the readers to \cite{Trottenberg:2000:MUL:374106} for a detailed introduction and briefly review the core idea here. 

\subsection{STFAS Algorithm \label{sec:Op}}
Our multigrid solver works on a hierarchy of grids in descending resolutions. In each STFAS iteration, it refines the solution $(v_i,\bar{p}_i,u_i,p_i)$ by reducing the residual $f(v_i,\bar{p}_i,u_i,p_i)$. Since different components of the residual can be reduced most effectively at different resolutions, the multigrid solver downsamples the residual to appropriate resolutions and then upsamples and combines their solutions. With properly defined operators introduced in this section, our multigrid algorithm can generally achieve a linear rate of error reduction, which is optimal in the asymptotic sense.

To adopt this idea to solve \prettyref{eq:KKTAO}, we introduce a hierarchy of spacetime grids $(v_i^h,\bar{p}_i^h,u_i^h,p_i^h)$, where $h$ is the cell size. We use semi-coarsening in spatial direction only where every coarser level doubles the cell size. We denote the coarser level as $(v_i^{2h},\bar{p}_i^{2h},u_i^{2h},p_i^{2h})$. We use the simple FAS-VCycle(2,2) iteration to solve the nonlinear system of equations: $f(v_i,\bar{p}_i,u_i,p_i)=\E{res}$. \changed{See \prettyref{Alg:STFAS} for details of the \AbbrAO solver.}
\setlength{\textfloatsep}{5pt}
\begin{algorithm}[h]
\caption{STFAS $\E{VCycle}(v_i^h,\bar{p}_i^h,u_i^h,p_i^h,\E{res}^h)$: This is used to solve the \AbbrAO subproblem. The algorithm is a standard FAS VCycle with 2 pre and post smoothing (\prettyref{ln:PRE}, \prettyref{ln:POST}) and 10 final smoothing (\prettyref{ln:FINAL}).}
\label{Alg:STFAS}
\begin{algorithmic}[1]
\Require A tentative solution $(v_i^h,\bar{p}_i^h,u_i^h,p_i^h)$
\Ensure Refined solution to $f(v_i^h,\bar{p}_i^h,u_i^h,p_i^h)=\E{res}^h$
\If{$h$ is coarsest}
\LineComment{Final smoothing for the coarsest level}
\For{$k=1,\cdots,10$}\label{ln:FINAL}
\State $\E{S}(v_i^h,\bar{p}_i^h,u_i^h,p_i^h)$
\EndFor
\Else
\LineComment{Pre smoothing}
\For{$k=1,2$}\label{ln:PRE}
\State $\E{S}(v_i^h,\bar{p}_i^h,u_i^h,p_i^h)$
\EndFor
\LineComment{Down-sampling}
\For{$t=v,\bar{p},u,p$ and $\forall i$}
\State $t_i^{2h}=\E{R}(t_i^{h})$
\State $t_i^{h}=t_i^{h}-\E{P}(t_i^{2h})$
\EndFor
\LineComment{Compute FAS residual by combining:}
\LineComment{1. the solution on coarse resolution}
\LineComment{2. the residual on fine resolution}
\State $\E{res}^{2h}=f(v_i^{2h},\bar{p}_i^{2h},u_i^{2h},p_i^{2h})$
\label{ln:FASRHS}
\State $\E{res}^{2h}=\E{res}^{2h}+\E{R}(\E{res}^h-f(v_i^h,\bar{p}_i^h,u_i^h,p_i^h))$
\LineComment{VCycle recursion}
\State $\E{VCycle}(v_i^{2h},\bar{p}_i^{2h},u_i^{2h},p_i^{2h},\E{res}^{2h})$
\LineComment{Up-sampling}
\For{$t=v,\bar{p},u,p$ and $\forall i$}
\State $t_i^{h}=t_i^{h}+\E{P}(t_i^{2h})$
\EndFor
\LineComment{Post smoothing}
\For{$k=1,2$}\label{ln:POST}
\State $\E{S}(v_i^h,\bar{p}_i^h,u_i^h,p_i^h)$
\EndFor
\EndIf
\end{algorithmic}
\end{algorithm}

\begin{figure}[h]
\begin{center}
\scalebox{0.9}{
\includegraphics[trim={2cm 7cm 2cm 6.5cm},clip,width=0.45\textwidth]{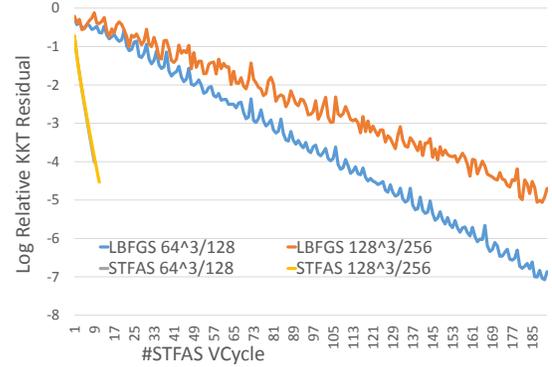}
}
\end{center}
\vspace{-15px}
\caption{\changed{\label{fig:Convergence} Convergence history of STFAS compared with that of the LBFGS optimizer, running on two grid resolutions and with a different number of timesteps (denoted as $n^d/N$). STFAS achieves a linear rate of error reduction independent of grid resolution and number of timesteps, as the two curves overlap.}}
\vspace{5px}
\end{figure}
The fast convergence of the geometric FAS relies on a proper definition of the three application-dependent operators: $\E{R},\E{P}$ and $\E{S}$. The restriction operator $\E{R}$ downsamples a fine grid solution to a coarser level for efficient error reduction, and the prolongation operator $\E{P}$ upsamples the coarse grid solution to correct the fine grid solution. We use simple trilinear interpolation for these two operators whether applied on scalar or vector fields. Finally, designing the smoothing operator $\E{S}$ is much more involved. $\E{S}$ should, by itself, be a cheap iterative solver for $f(v_i,\bar{p}_i,u_i,p_i)=\E{res}$. Compared with previous works such as \cite{chentanez2011real} where multigrid is used for solving the pressure field $p_i$ only, we are faced with two new challenges. First, since we are solving the primal as well as dual variables, which gives a saddle point problem, the Hessian matrix is not positive definite in the spatial domain, so that a Jacobi or Gauss-Seidel (GS) solver does not work. Second, we are not coarsening in the temporal domain, so the temporal correlation must be considered in the smoothing operator. 

Our solution is to consider the primal and dual variables at the same time using the Symmetric Coupled Gauss-Seidel (SCGS) smoothing operator \cite{vanka1983fully}. SCGS smoothing is a primal-dual variant of GS. In our case, where all the variables are stored in a staggered grid, SCGS smoothing considers one cell at a time. It solves the primal variables $v_i,u_i$ stored on the $6$ cell faces as well as the dual variables $p_i,\bar{p_i}$ stored in the cell center at the same time by solving a small $14\times14$ linear problem ($10\times10$ in 2D). Like red-back-GS smoothing, we can parallelize SCGS smoothing using the $8$-color tagging (see \prettyref{fig:STFAS}). 

The above SCGS solver only considers one timestep at a time. To address the second problem of temporal correlation, we augment the SCGS solver with the temporal domain. We solve the $14$ variables associated with a single cell across all the timesteps at once. Although this involves solving a large $14N\times 14N$ linear system for each cell, the left hand side of the linear system is a block tridiagonal matrix so that we can solve the system in $\mathcal{O}(N)$. Indeed, the Jacobian matrix of $f$ takes the following form:
\begin{align}
\label{eq:HESS}
\resizebox{0.7\hsize}{!}{
$\FPP{f}{v_i,\bar{p}_i,u_i,p_i}=
\left(\begin{array}{cc|cc|ccc}
\frac{K}{r}I   & \nabla & \FPP{u_0}{v_0}^T &        &                &        & \\ 
\nabla^T       &        &                  &        &                &        & \\
\hline
\FPP{u_0}{v_0} &        & -I               & \nabla & \FPP{u_0}{v_1} &        & \\ 
               &        & \nabla^T         &        &                &        & \\
\hline
               &        & \FPP{u_0}{v_1}^T &        & \frac{K}{r}I   & \nabla & \\ 
               &        &                  &        & \nabla^T       &        & \\
               &        &                  &        &                &        & \ddots \\
\end{array}\right)$,
}
\end{align}
where the size of each block is $5\times5$ in 2D and $7\times7$ in 3D. Due to this linear time solvability, the optimal multigrid performance is still linear in the number of spatial-temporal variables. The average convergence history for our multigrid solver is compared with a conventional LBFGS algorithm \cite{mcnamara2004fluid} in \prettyref{fig:Convergence}. Our algorithm achieves a stable linear rate of error reduction independent of both the grid resolution and the number of timesteps.

\subsection{ADMM Outer Loop}
Equipped with solvers for the two subproblems, \changed{we present our ADMM outer loop in \prettyref{Alg:outter}}. We find it difficult for either \prettyref{eq:KKTPO} or a quasi-Newton method solving the \AbbrPO subproblem to converge to an arbitrarily small residual due to the non-smooth nature of the operator $\ARho{\bullet}{\bullet}$. Both algorithms decrease the objective function in the first few iterations and then wander around the optimal solution. In view of this, we run \prettyref{eq:KKTPO} (\prettyref{sec:PO}) for a fixed number of iterations before moving on to the \AbbrAO subproblem (\prettyref{sec:AO}) so that each ADMM iteration has $\mathcal{O}(n^dN)$ complexity and is linear in the number of spacetime variables. Finally our stopping criterion for the \AbbrAO subproblem is that the residual $\|f\|_{\infty}<\epsilon_{STFAS}$. Our stopping criterion for the ADMM outer loop is that the maximal visual difference, the largest difference of the density field over all the timesteps, generated by two consecutive ADMM iterations should be smaller than $\epsilon_{ADMM}$.
\setlength{\textfloatsep}{5pt}
\begin{algorithm}[h]
\caption{ADMM Outer Loop}
\label{Alg:outter}
\begin{algorithmic}[1]
\Require Parameters $K,r,\rho_i^*,\epsilon_{STFAS},\epsilon_{ADMM}$
\Ensure Optimized velocity fields $v_i$ and density fields $\rho_i$
\For{$i=0,\cdots,N$}
\State Set $v_i=0$
\State Set $\rho_i^{last}=\rho_i$
\EndFor
\While {true}   \label{ln:ADMMOuter}
\LineComment{Solve the \AbbrPO subproblem}
\State Run \prettyref{Alg:Passive} for a fixed number of iterations \label{ln:AO}
\LineComment{Solve the \AbbrAO subproblem}
\While {$f(v_i,\bar{p}_i,u_i,p_i)>\epsilon_{STFAS}$}    \label{ln:NSO}
\State \prettyref{Alg:STFAS}
\EndWhile
\LineComment{Stopping criterion}
\If{$\E{max}_i\|\rho_i^{last}-\rho_i\|_{\infty}<\epsilon_{ADMM}$}
\State Return $v_i,\rho_i$
\EndIf
\For{$i=0,\cdots,N$}
\State Set $\rho_i^{last}=\rho_i$
\LineComment{Update augmented Lagrangian multiplier}
\label{ln:AugLag}
\State Set $\lambda_i=\lambda_i+K\beta(v_i-v_i^*)$
\EndFor
\EndWhile
\end{algorithmic}
\end{algorithm}

\section{Results and Analysis} \label{sec:results}
\begin{figure}[h]
\vspace{-15px}
\begin{center}
\includegraphics[trim={2cm 8cm 1.5cm 9cm},clip,width=0.23\textwidth]{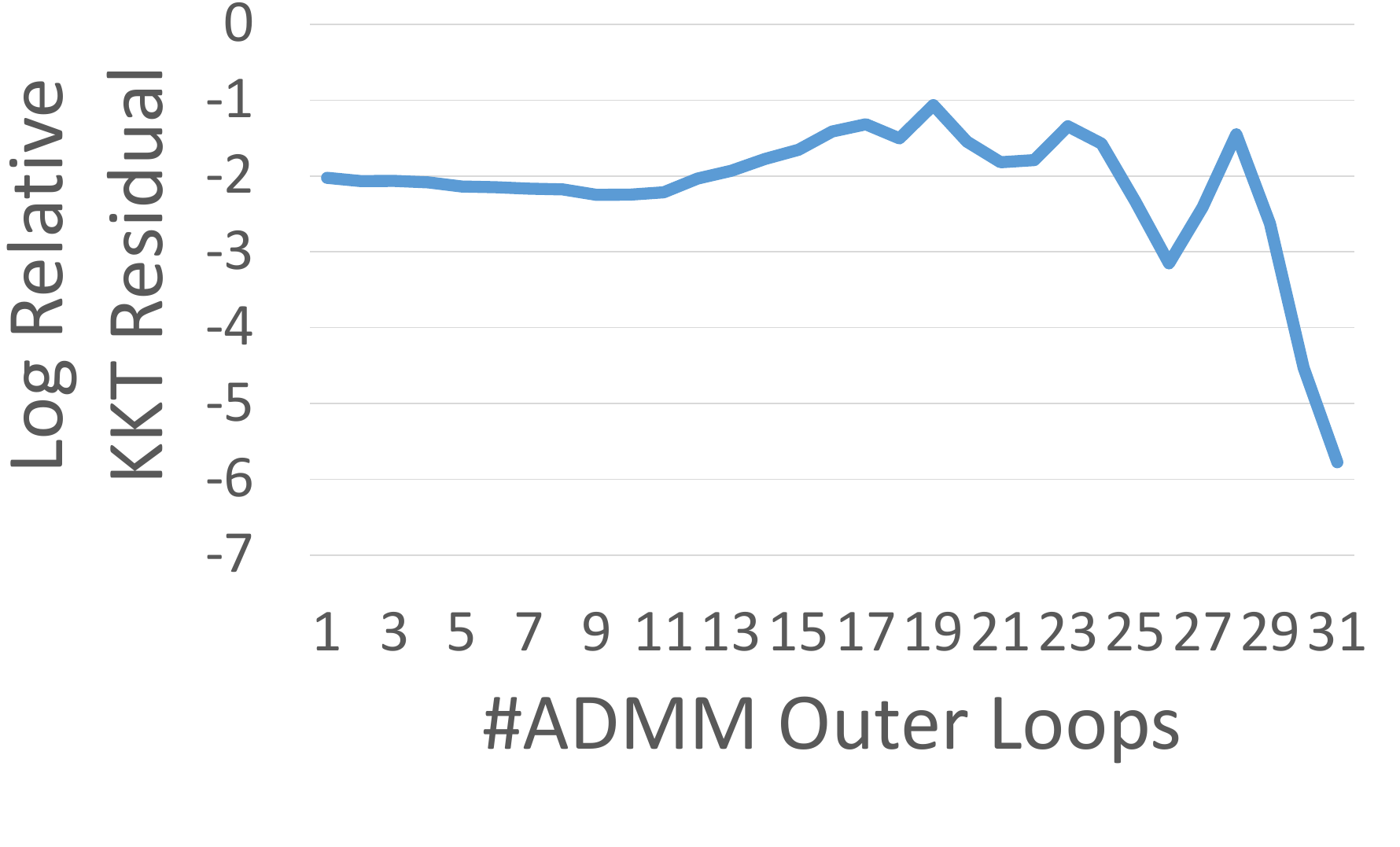}
\includegraphics[trim={2cm 2cm 1.5cm 2cm},clip,width=0.23\textwidth]{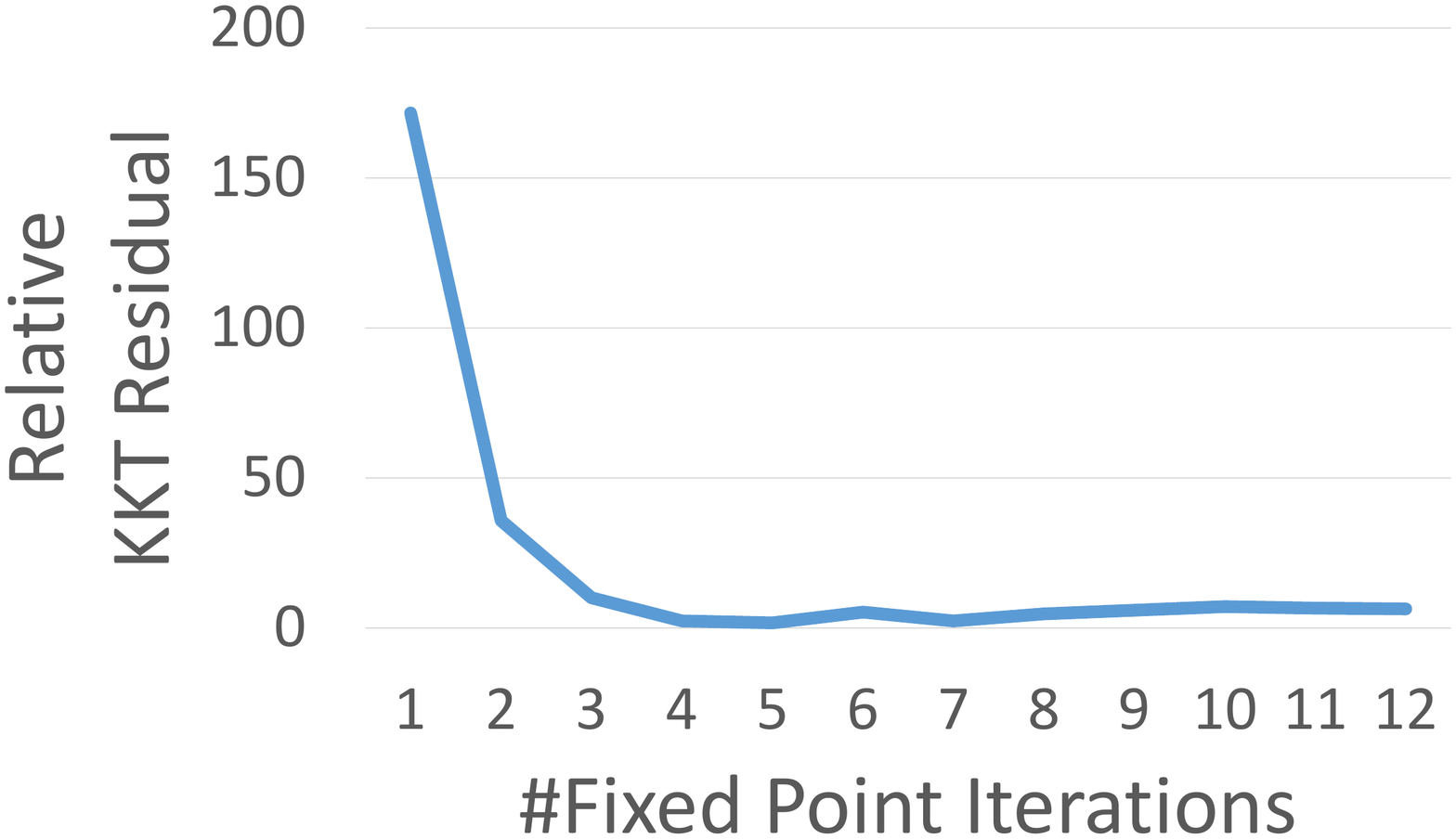}
\end{center}
\vspace{-20px}
\caption{\label{fig:ConvergenceADMM} We profile the convergence history of the example \prettyref{fig:Teaser}. We plot the logarithm of relative KKT residual of the optimized velocity field after each ADMM loop (left); and the absolute residual of the \AbbrPO subproblem's KKT conditions after each iteration of \prettyref{eq:KKTPO} (right).}
\vspace{-10px}
\end{figure}

\begin{figure}[h]
\begin{center}
\includegraphics[trim={0cm 0cm 0cm 0cm},clip,width=0.47\textwidth]{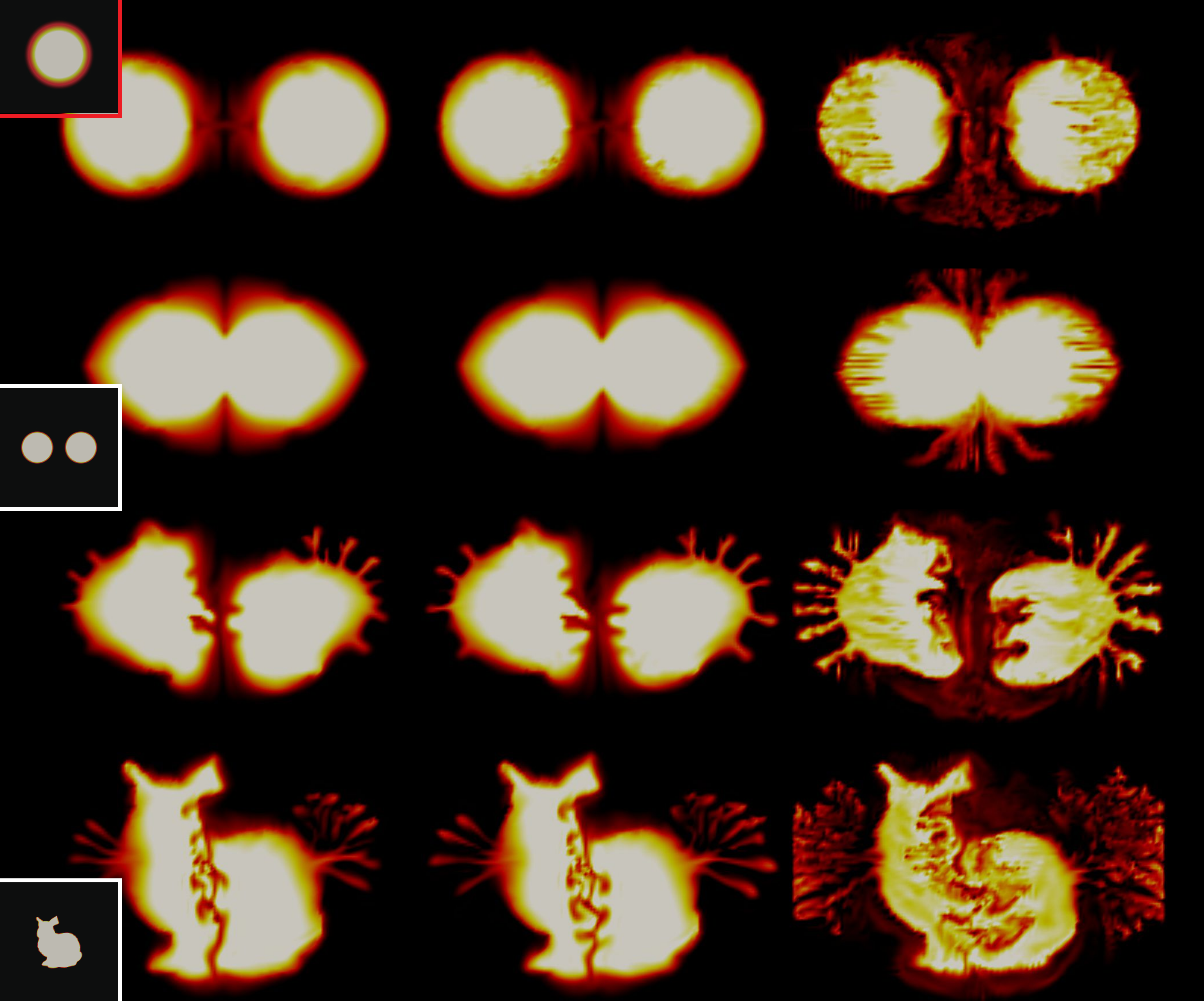}
\end{center}
\caption{\changed{\label{fig:Reg} For this animation, we match the circle (red) first to two smaller circles and then to a bunny (we show frames $20,40,60,80$ from top to bottom). The resolution is $128^2/80$, and we test three different values of ghost force regularization $r=10^{2,3,4}$ (from left to right). More smoke-like behaviors are generated as we increase $r$.}}
\end{figure}

\begin{figure}[h]
\vspace{5px}
\begin{center}
\includegraphics[trim={0cm 0cm 0cm 0cm},clip,width=0.48\textwidth]{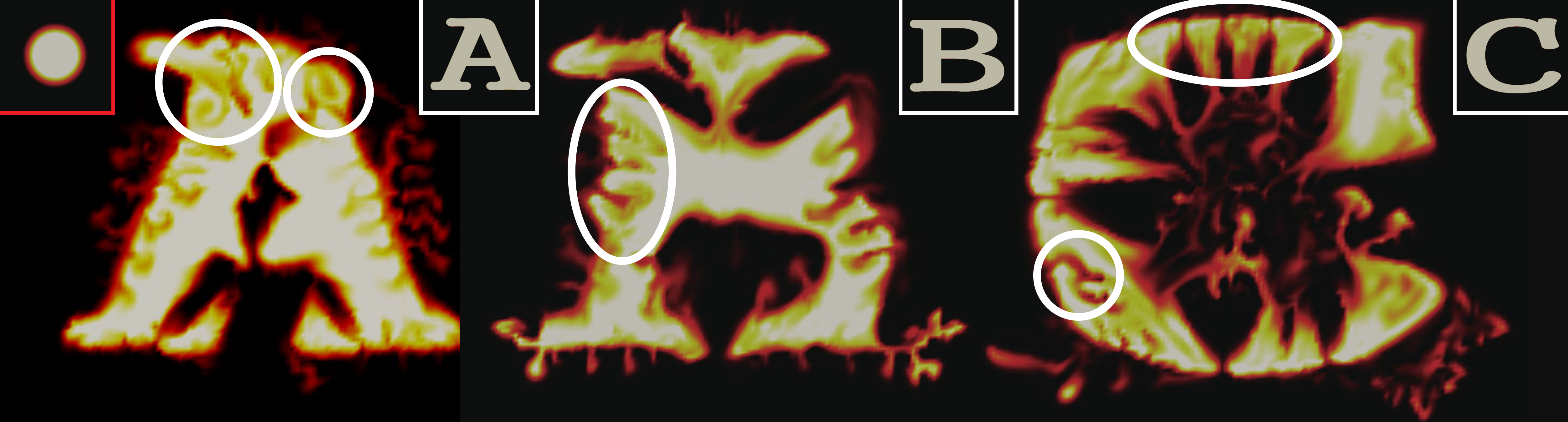}
\end{center}
\caption{\changed{\label{fig:Reg2} In this example, we deform a sphere into letter ``A'', then letter ``B'' and finally letter ``C''. For such complex deformation, it is advantageous to allow every velocity component to be optimized. So that a lot of fine-scale details can be generated as illustrated in the white circles.}}
\vspace{5px}
\end{figure}

\setlength{\tabcolsep}{5pt}
\begin{wraptable}{l}{4.0cm}
\vspace{-15px}
\begin{center}
\begin{tabular}{lc}
Name & Value   \\
\hline  \\
$\Delta t$ & $0.4\sim2.0s$  \\
$K$ & $10^3$  \\
$r$ & $10^{2\sim4}$   \\
$\beta$ for updating $\lambda_i$ & $1$    \\
\#\prettyref{eq:KKTPO} & $2$ \\
$\epsilon_{STFAS}$ & $10^{-5}$  \\
$\epsilon_{ADMM}$ & $\frac{\rho_{max}}{100}$   \\
\end{tabular}
\end{center}
\caption{\label{table:param} Parameters.}
\vspace{0px}
\end{wraptable}
\TE{Parameter Choice:} We use the same set of parameters listed in \prettyref{table:param} for all experiments, where $\rho_{max}$ is the maximal density magnitude at the initial frame. Under this setting, the convergence history of the ADMM outer loop of our first example \prettyref{fig:Teaser} is illustrated in \prettyref{fig:ConvergenceADMM}. In our experiments, the ADMM algorithm always converges in fewer than 50 iterations. Further, running only 2 iterations of \prettyref{eq:KKTPO} in each ADMM loop will not deteriorate the performance. In fact, according to the averaged convergence history of the \AbbrPO subproblem illustrated in \prettyref{fig:ConvergenceADMM}, the fixed point iteration \prettyref{eq:KKTPO} usually converges in the first 4 iterations before it wanders around a local minimum. After fine tuning, we found that 2 iterations lead to the best overall performance. In this case, the overhead of solving the \AbbrPO subproblem is marginal compared with the overhead of solving the \AbbrAO subproblem. \changed{Finally, unlike fluid simulation, the performance of spacetime optimization doesn't depend on the timestep size due to our robust advection operator (\prettyref{eq:Adv}). When we increase the timestep size from $0.4s$ to $2s$ for the examples in \prettyref{fig:Reg} and \prettyref{fig:Reg2}, which is extremely large, our algorithm's convergence behavior is about the same.}

\begin{table*}[t]
\setlength{\tabcolsep}{8pt}
\begin{center}
\scalebox{0.8}{
\begin{tabular}{lccccccc}
\toprule
Example$(n^d/N)$ & Boundary & \#ADMM & Avg. \AbbrPO(s) & Avg. \AbbrAO(s) & Total(hr) & Memory(Gb) & Total LBFGS(hr)	\\
\midrule
Letters FLUID$(128^2/40,r=10^3)$ & Neumann & 13 & 10 & 60  & 0.25 & 0.06 & 4  \\
Circle Bunny$(128^2/80, r=10^2)$ & Neumann & 25 & 20 & 130 & 1.04 & 0.2  & 12  \\
Circle Bunny$(128^2/80, r=10^3)$ & Neumann & 37 & 20 & 220 & 2.46 & 0.2  & 15  \\
Circle Bunny$(128^2/80, r=10^4)$ & Neumann & 43 & 20 & 218 & 2.84 & 0.2  & 16  \\
Letters ABC$(128^2/60, r=10^4)$ & Neumann & 33 & 16 & 179 & 1.78 & 0.15  & 14  \\
Sphere Armadillo Bunny$(64^3/40, r=10^3)$ & Neumann & 17 & 103 & 1341 & 6.81 & 1.34 & N/A   \\
Varying Genus$(64^2\times 32/40, r=10^3)$ & Periodic & 20 & 82 & 840 & 5.12 & 0.67 & N/A  \\
Human Mocap$(64^2\times 128/60, r=10^3)$ & Periodic & 5 & 1437 & 3534 & 6.9 & 4.0 & N/A  \\
Moving Sphere$(64^3/60, r=10^2)$ & Neumann & 17 & 630 & 1792 & 11.43 & 2.2 & N/A  \\
Moving Sphere$(64^3/60, r=10^3)$ & Neumann & 22 & 630 & 1978 & 15.93 & 2.2  & N/A  \\
\bottomrule
\end{tabular}
}
\end{center}
\caption{\changed{\label{table:Perf} Memory and computational overhead for all the benchmarks. From left to right: name of example (resolution parameters); the spatial boundary condition; number of outer ADMM iterations; average time spent on each \AbbrPO subproblem; average time spent on each \AbbrAO subproblem; total time until convergence using our algorithm; memory overhead; total time until convergence using LBFGS. By comparing the three ``Circle Bunny'' examples, we can see that the number of ADMM outer loops is roughly linear to $log_{10}(r)$. More ADMM outer loops are needed, if more fluid-like behaviors are desired. And the computational cost of each ADMM outer loop is roughly linear in the number of timesteps. This can be verified by comparing the ``Letters FLUID'' and the ``Circle Bunny'' example.}}
\vspace{5px}
\end{table*}


\TE{Benchmarks:} To demonstrate the efficiency and robustness of our algorithm, we used 7 benchmark problems that vary in their grid resolution, number of timesteps, and number of keyframes. The memory overhead and computational overhead are summarized in \prettyref{table:Perf}. All of the results are generated on a desktop PC with an i7-4790 8-core CPU 3.6GHz and 12GB of memory. We use OpenMP for multithread parallelization.

Our first example is five controlled animations matching a circle to the letters ``FLUID''. Compared with \cite{treuille2003keyframe}, which uses a relatively small set of control force templates to reduce the search space of control forces, we allow control on every velocity component so that the matching to keyframe is almost exact. After the keyframe, we remove the control force, and rich smoke details are generated by pure simulation as illustrated in \prettyref{fig:Teaser}. However, in the controlled phase of \prettyref{fig:Teaser}, this example seems ``too much controlled'', meaning that most smoke-like behaviors are lost. This effect has also been noticed in \cite{treuille2003keyframe}. However, unlike their method, in which the number of templates needs to be carefully tuned to recover such behavior, we can simply adjust the regularization $r$ in our system to balance matching exactness and the amount of smoke-like behaviors. In \prettyref{fig:Reg}, we generated three animations with two keyframes: first two circles and then a bunny, using $r=10^{2,3,4}$ respectively. These animations are also shown in the video. Our algorithm is robust to a wide range of parameter choices. But more iterations are needed for the multigrid to converge for a larger $r$ as shown in \prettyref{table:Perf}. \changed{Finally, since we allow every velocity component to be optimized, the resulting animation exhibits lots of small-scale details as indicated in \prettyref{fig:Reg2}, which is not possible with the small set of force templates used in \cite{treuille2003keyframe}.}

\begin{figure*}[t]
\begin{center}
\includegraphics[width=0.98\textwidth]{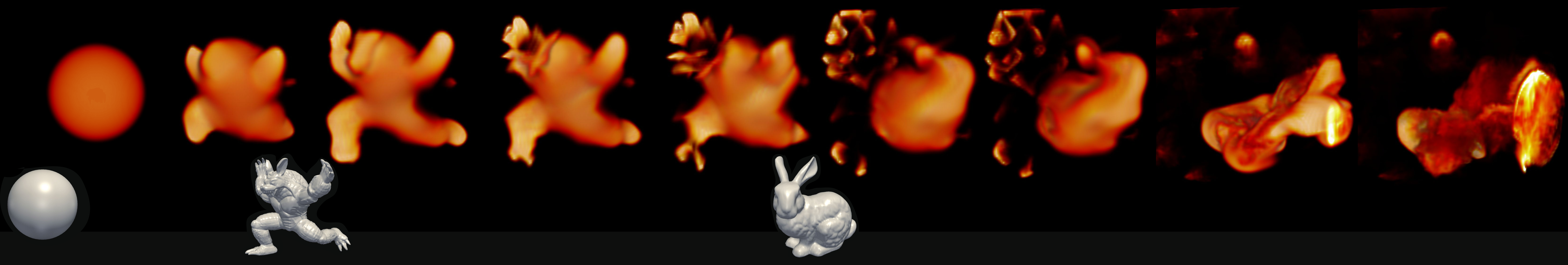}
\end{center}
\caption{\changed{\label{fig:ArmadilloBunny} 3D smoke control example of deforming a sphere first to an armadillo and then to a bunny. This example runs at the resolution of $64^3$ with $40$ timesteps. The optimization can be accomplished in $7$hr.}}
\vspace{-5px}

\end{figure*}
\begin{figure*}[t]
\begin{center}
\includegraphics[width=0.9\textwidth]{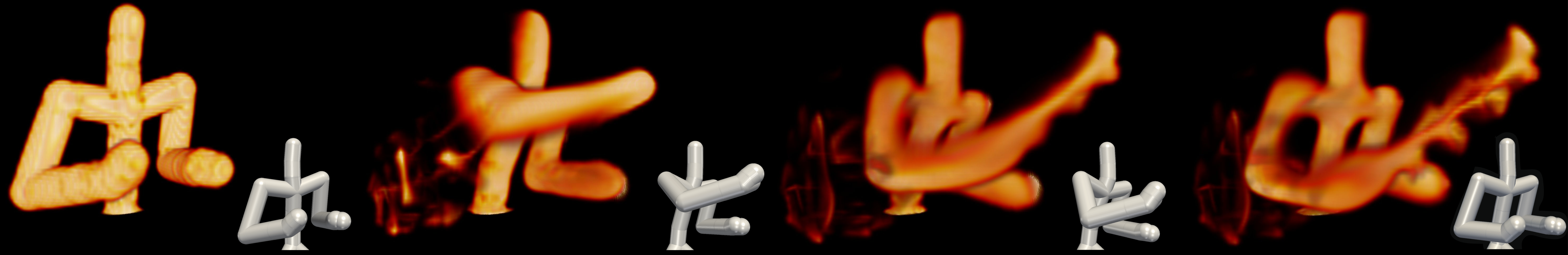}
\end{center}
\caption{\changed{\label{fig:Punch} We generate the famous example of tracking smoke with a dense sequence of keyframes, which comes from human motion capture data. Our algorithm converges and generates rich smoke drags within 5 ADMM iterations.}}
\vspace{-5px}
\end{figure*}

\begin{figure*}[t]
\begin{center}
\includegraphics[width=0.98\textwidth]{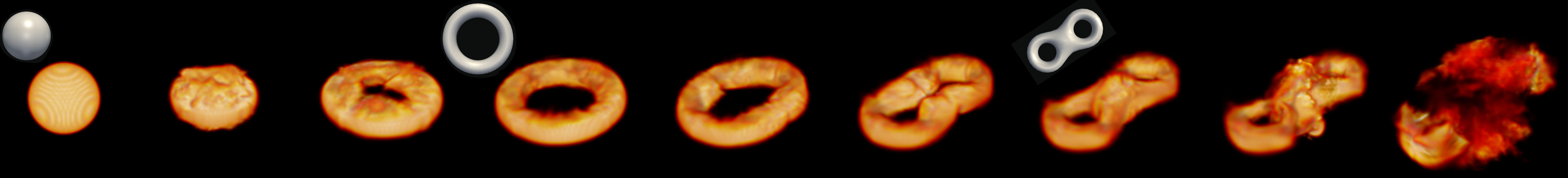}
\end{center}
\vspace{-5px}
\caption{\changed{\label{fig:GV} Example of smoke control where the keyframes have varying genera. The initial frame is a sphere (genus 0). The first keyframe located at frame 20 is a torus (genus 1) and the second keyframe located at frame 40 is the shape eight (genus 2). The resolution is $64^2\times 32/40$ and the overall optimization takes 5hr with $r=10^3$.}}
\vspace{-10px}
\end{figure*}

\begin{figure*}[t]
\begin{center}
\includegraphics[width=0.9\textwidth]{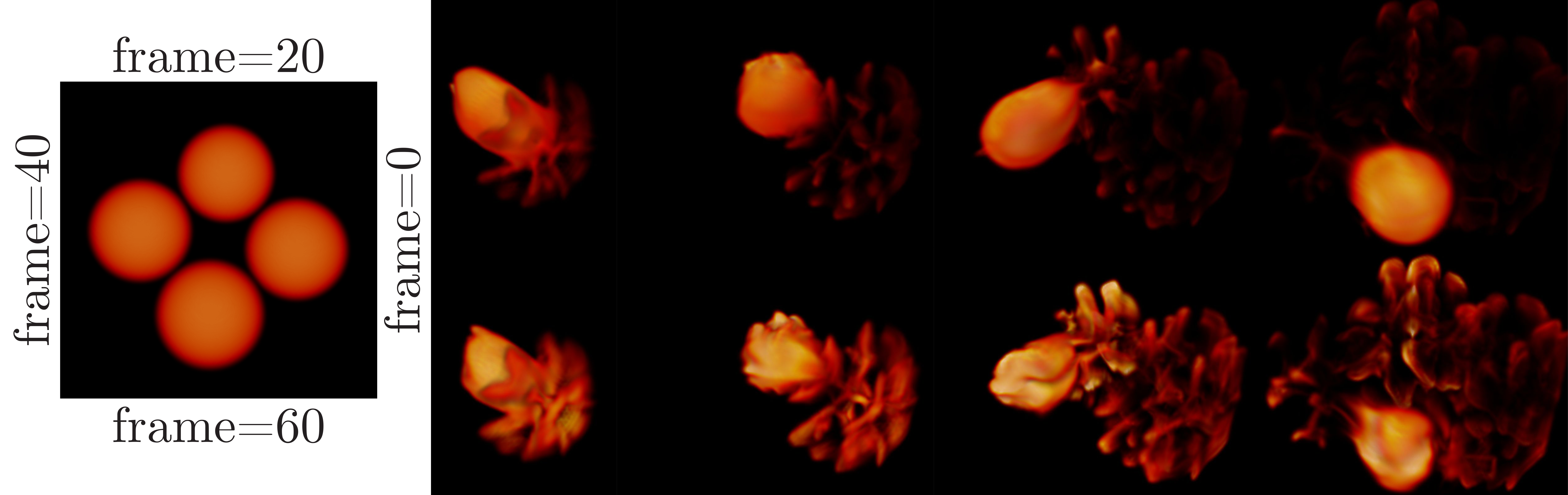}
\end{center}
\vspace{-5px}
\caption{\changed{\label{fig:SphereMove} A moving smoke sphere guided by the 3 keyframes (left). We experimented with $r=10^3$ (top) and $r=10^4$ (bottom). Larger regularization results in more wake flow behind moving smoke bodies. The same effect can be observed in \prettyref{fig:Reg}.}}
\vspace{-10px}
\end{figure*}

In addition to these 2D examples, we also tested our algorithm on some 3D benchmarks. Our first example is shown in \prettyref{fig:ArmadilloBunny} and runs at a resolution of $64^3/40$. We use two keyframes at frame $20$ and $40$, and the overall optimization takes about 7 hours. In our second example, shown in \prettyref{fig:Punch}, we try to track the smoke with a dense sequence of keyframes from the motion capture data of a human performing a punch action. Such an example is considered the most widely used benchmarks for PD-type controllers such as \cite{shi2005taming}. With such strong and dense guidance, our algorithm converges very quickly, within $5$ iterations. \changed{Our third example (\prettyref{fig:SphereMove}) highlights the effect of regularization coefficient $r$ in 3D. Like our 2D counterpart \prettyref{fig:Reg}, larger $r$ usually results in more wake flow behind moving smoke bodies. Finally, we evaluated our algorithm on a benchmark with keyframe shapes of varying genera. As illustrated in \prettyref{fig:GV}, the initial smoke shape has genus zero, but we use two keyframes, where the smoke shapes have genus one and two. Our algorithm can handle such complex cases.}

\TE{Comparison with LBFGS}: We compared our algorithm with a gradient-based quasi-Newton optimizer. Specifically, we use LBFGS method \cite{nocedal2006numerical}. Such method approximates the Hessian using a history of gradients calculated by past iterations. We set the history size to be 8, which is typical. We use same stopping criteria for both LBFGS and our method. Under this setting, we compared the performance of LBFGS and the ADMM solver on two of our 2D examples: \prettyref{fig:Teaser} and \prettyref{fig:Reg}. For the example of letter matching in \prettyref{fig:Teaser}, LBFGS algorithm takes 4hr and 71 iterations to converge. While for the example of changing regularization in \prettyref{fig:Reg}, LBFGS algorithm takes 12hr and 152 iterations at $r=10^2$, 15hr and 170 iterations at $r=10^3$, and 16hr and 212 iterations at $r=10^4$. Therefore, our algorithm is approximately an order of magnitude faster than a typical implementation of LBFGS. 

The speedup over LBFGS optimizer occurs for two reasons. First, we break the problem up into the \AbbrPO subproblem and the \AbbrAO subproblem, that have sharply different properties. The \AbbrPO subproblem is nonsmooth while the \AbbrAO subproblem is not. In practice, neither our fixed point iteration scheme in \prettyref{eq:KKTPO} nor the LBFGS algorithm can efficiently solve \AbbrPO to arbitrarily small KKT residual. Without such decomposition, it takes a very long time to solve the overall optimization problem by taking a lot of iterations. The second reason is the use of warm-started STFAS solver for the \AbbrAO subproblem. Note that LBFGS algorithm not only takes more iterations, but each iteration is also more expensive. This is mainly because of the repeated gradient evaluation in each LBFGS iteration, where each evaluation runs the adjoint method with a cost equivalent to two passes of fluid resimulation.

\begin{changedBlk}
\TE{Comparison with PD Controller:} We also compared our method with simple tracker type controllers such as PD controller \cite{fattal2004target}. To drive the fluid body towards a target keyframe shape using heuristic ghost forces, PD controllers result in much lower overhead in terms of fluid resimulation, as compared to our approach based on optimal controllers. In contrast, optimal controllers provide better flexibility and robust solutions as compared to PD controllers. A PD controller tends to be very sensitive to the parameters of the ghost force. Moreover, its performance also depends on the non-physical gathering term to generate plausible results (see \prettyref{fig:PDSensitive}). On the other hand, an optimal controller can achieve exact keyframe timing, which may not be possible using a PD controller, as shown in \prettyref{fig:PDSensitive}. Moreover, an optimal controller can easily balance between the exactness of keyframe matching and the amount of fluid-like behavior based on a single tuning parameter $r$ (see \prettyref{fig:Reg}).
\end{changedBlk}

\begin{changedBlk}
\TE{Memory Overhead:} Since fluid control problems usually have a high memory overhead, we derive here an analytical upper bound of the memory consumption $M(n,d,N)$:
\begin{eqnarray*}
\resizebox{0.99\hsize}{!}{
$M(n,d,N)\sim\left[(n^d)*(1+d)*2*2\right]*\left[1+\frac{1}{2}+\frac{1}{4}\cdots\right]*N=8n^d(1+d)N,$}
\end{eqnarray*}
where $n$ is the grid resolution, $d$ is the dimension, and $N$ is the number of timesteps. To derive this bound, note that we can reuse the memory consumed by \prettyref{Alg:STFAS} in \prettyref{Alg:Passive}, and \prettyref{Alg:STFAS} always consumes more memory than \prettyref{Alg:Passive}, so that we only consider the memory overhead of \prettyref{Alg:STFAS}. The first term $n^d*(1+d)$ is the number of variables needed for storing a pair of pressure and velocity fields. This number is doubled because we need to store $u_i,\bar{p}_i$ in addition to $v_i,p_i$ at each timestep. We double it again because we need additional memory for storing $\E{res}$ in STFAS. Finally, the power series is due to the hierarchy of grids. At first observation, this memory overhead is higher than \cite{treuille2003keyframe,mcnamara2004fluid} since we require additional memory for storing the dual variables at multiple resolutions. However, due to the quasi-Newton method involved in their approach, additional memories are needed to store a set of $L$ gradients to approximate the inverse of the Hessian matrix. $L$ is usually $5\sim10$, leading to the following upper bound:
\begin{eqnarray*}
M_{LBFGS}(n,d,N)\sim\left[(n^d)*(1+d)\right]*L*N=Ln^d(1+d)N.
\end{eqnarray*}
In our benchmarks, the memory overheads of our ADMM and LBFGS solvers are comparable. 
\end{changedBlk}

\begin{changedBlk}
\TE{Convergence Analysis:} Here we analyze the convergence of our approach and propose some modifications to guarantee convergence of iterations used in \prettyref{Alg:outter}.

For our \AbbrPO solver (\prettyref{ln:AO} of \prettyref{Alg:outter}), we observe that it can be difficult for \prettyref{Alg:Passive} to converge to an arbitrarily small KKT residual in each loop of \prettyref{Alg:outter}. Although our choice is to perform a fixed number of iterations of \prettyref{Alg:Passive}, one could also use a simple strategy that can guarantee that function value decreases by blending a new solution with the previous solution and tuning the blending factor in a way similar to the line search algorithm. This modification has low computational overhead since one doesn't need to apply the costly solenoidal projection operator $\E{Q}$ again after the blending, as the sum of two solenoidal vector fields is still solenoidal. In our benchmarks, this strategy leads to a convergent algorithm with low overhead, but the error reduction rate after the first few iterations can be slow.

\begin{figure}[h]
\begin{center}
\includegraphics[trim={2cm 7cm 1cm 7cm},clip,width=0.49\textwidth]{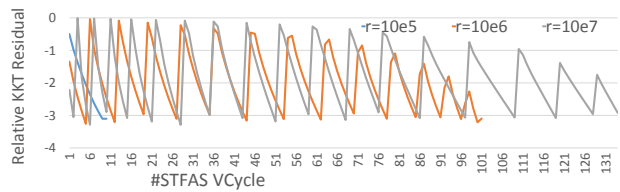}
\end{center}
\vspace{-15px}
\caption{\changed{\label{fig:convLargeR} Convergence history of the \AbbrAO solver in the Circle Bunny example. When the regularization coefficient $r$ is extremely large, we have to treat the FAS-Vcycle as a subproblem solver of the LM algorithm. In each subproblem solve, the convergence rate of FAS-VCycle is still linear.}}
\end{figure}
The same analysis can also be used for the \AbbrAO solver (\prettyref{ln:NSO} of \prettyref{Alg:outter}) as well. To ensure convergence of \prettyref{Alg:STFAS}, we could add a perturbation to the penalty coefficient $K$ in the Hessian matrix \prettyref{eq:HESS}. Note that as $K\rightarrow\infty$, $v_i\rightarrow v_i^*$. Therefore, this strategy essentially makes \prettyref{Alg:STFAS} the subproblem solver for the Levenberg-Marquardt algorithm \cite{nocedal2006numerical}, which in turn guarantees convergence. As illustrated in \prettyref{fig:convLargeR}, Levenberg-Marquardt modification can be necessary when one uses larger regularization $r$, because we observe that the convergence rate decreases as $r$ increases.

Finally, for the ADMM outer loop (\prettyref{ln:ADMMOuter} of \prettyref{Alg:outter}), current analysis of its convergence relies on strong assumptions of its objective function, such as global convexity. However, ADMM, as a variant of the Augmented Lagrangian solver \cite{nocedal2006numerical}, is guaranteed to converge by falling back to a standard Augmented Lagrangian solver. Specifically, one can run \prettyref{Alg:outter} without applying \prettyref{ln:AugLag} until the decrease in function value is lower than some threshold.

We have applied the above modifications for \prettyref{ln:AO} and \prettyref{ln:NSO} of \prettyref{Alg:outter}, which then takes a slightly more complex form. However, the introduced computational overhead is marginal. 
\end{changedBlk}
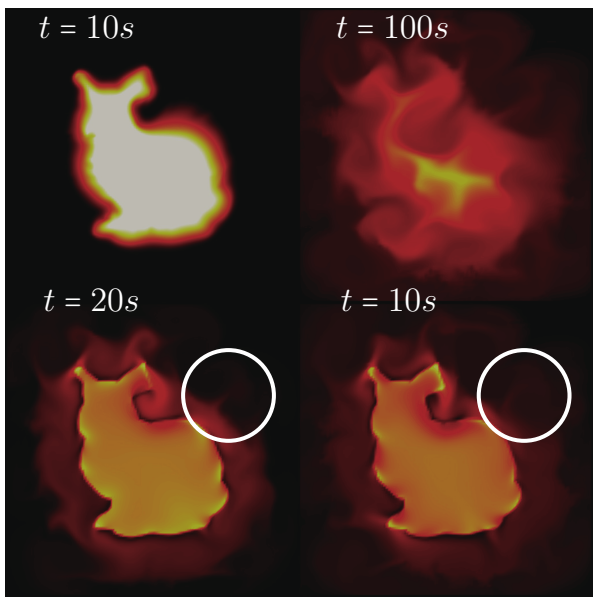
\begin{figure}
\begin{center}
\scalebox{0.9}{
\includegraphics[trim={0cm 0cm 0cm 0cm},clip,width=0.49\textwidth]{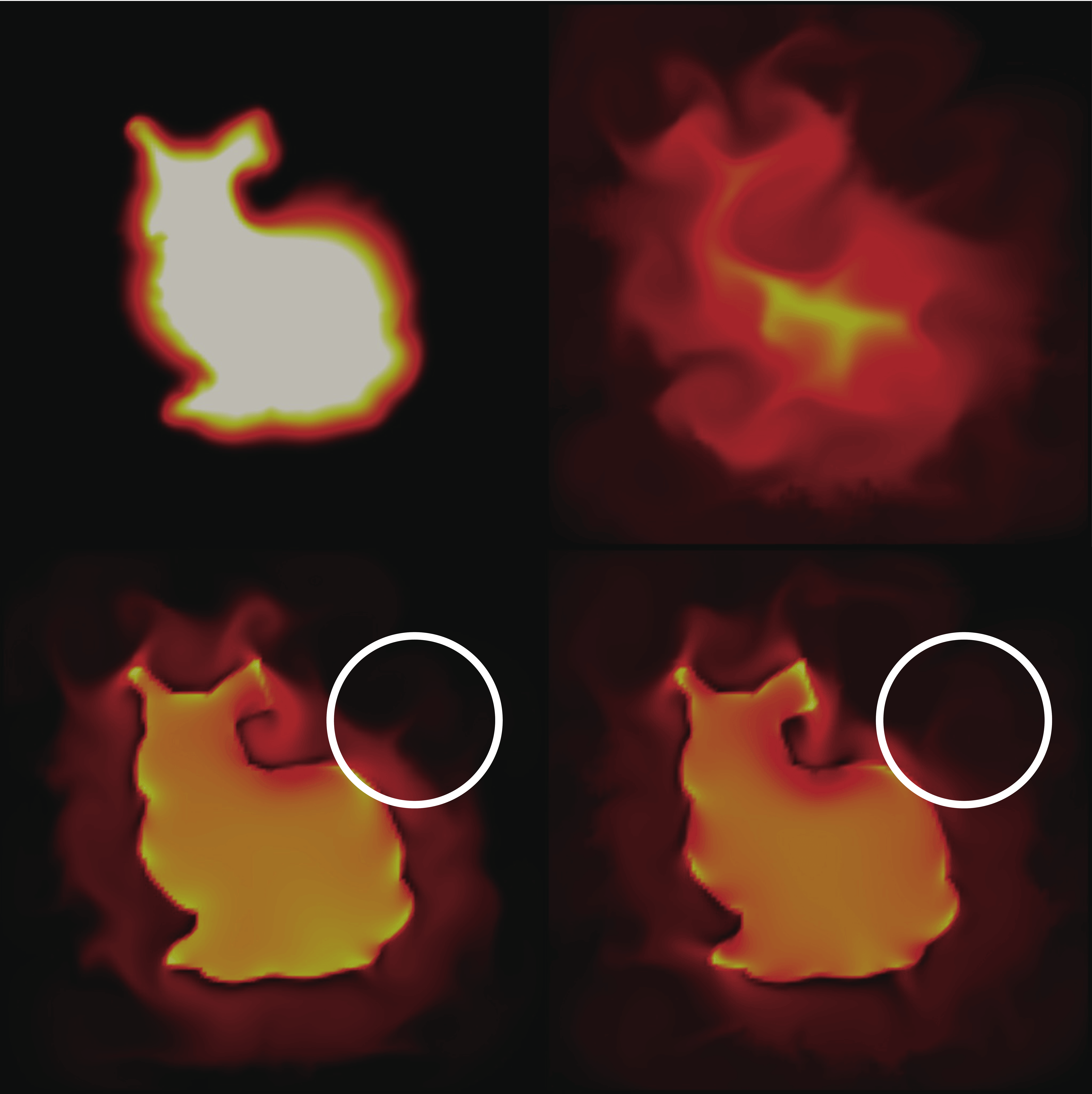}
\textcolor{white}{\put(-235,235){\Large$t=10s$}}
\textcolor{white}{\put(-235,120){\Large$t=20s$}}
\textcolor{white}{\put(-114,235){\Large$t=100s$}}
\textcolor{white}{\put(-114,120){\Large$t=10s$}}
}
\end{center}
\vspace{-5px}
\caption{\changed{\label{fig:PDSensitive} In order to deform the smoke circle into the bunny, the optimal controller achieves exact and reliable keyframe timing and produces a smooth matched shape (top left). But PD-controller requires careful tuning of ghost force coefficient to achieve such exact timing. In addition, there is still some wandering smoke (white circle) around the keyframe shape (bottom right). If one uses larger ghost force coefficient, the smoke vibrates around the keyframe and a stable matching occurs $10$s later (bottom left), with more wandering smokes (white circle). Moreover, PD-controller relies on the non-physical gathering term to generate plausible results. Without this gathering term, the keyframe is not matched, regardless of the length of the animation (top right).}}
\end{figure}

\section{Conclusion and Limitations}
In our work, we present a new algorithm for the optimal control of smoke animation. Our algorithm finds the stationary point of the KKT conditions, solving for both primal and dual variables. Our key idea is to refine primal as well as dual variables in a warm-started manner, without requiring them to satisfy the Navier-Stokes equations exactly in each iteration. We tested our approach on several benchmarks and a wide range of parameter choices. The results show that our method can robustly find the locally optimal control forces while achieving an order of magnitude speedup over the gradient-based optimizer, which performs fluid resimulation in each gradient evaluation.

On the downside, our method severely relies on the spatial structure and the staggered grid discretization of the Navier-Stokes equations. This imposes a major restriction to the application of our techniques. Nevertheless, generalizing our idea to other fluid discretization is still possible. For example, our method can be used with a fluid solver discretized on a general tetrahedron mesh such as \cite{Chentanez:2007:LSL,pavlov2011structure} since the KKT conditions are invariant under different discretizations, and the three operators to define STFAS stay valid. \changed{On the other hand, generalizing our method to free-surface flow or to handle internal boundary conditions can be non-trivial. The distance metric $C_i$ in \prettyref{eq:PO} needs to be modified to make it aware of the boundaries, e.g., Euclidean distances should be replaced with Geodesic distances. However, modifying the \AbbrAO solver to handle the boundaries can be relatively straightforward. This is because our multigrid formulation is the same as a conventional multigrid formulation in spatial domain, using simple trilinear prolongation and restriction operators. Therefore, existing works on boundary aware multigrid such as \cite{chentanez2011real} can also be applied to our spacetime formulation.}

In addition, unlike \cite{treuille2003keyframe,mcnamara2004fluid}, which use a set of template ghost force bases to reduce the search space, our method allows every velocity component to be optimized. This choice is application dependent. For matching smoke to detailed keyframes with lots of high frequency features, our formulation can be useful. However, using a reduced set of template ghost forces could help to avoid the popping artifacts illustrated in \prettyref{fig:Popping}, and at the same time it allows more user control over the applied control force patterns. \changed{For example, the use of vortex force templates encourages more swirly motions in the controlled animations.} Combining the control force templates with our formulation is considered as future work.

\changed{In terms of computational overhead, since our optimal controller always solves the spacetime optimization by considering all the timesteps, it is much slower than a simple PD controller which considers one timestep at a time. In order to reduce runtime cost, we can use a larger timestep size to reduce the number of timesteps. Our novel advection operator \prettyref{eq:Adv} can robustly handle this setting. Also, we can lower the spatial resolution in the control phase and then use smoke upsampling methods such as \cite{nielsen2011guide} to generate a high quality animation.}

\changed{In addition, further accelerations to our method are still possible. For example, parallelization of our algorithm in a distributed environment is straightforward. Indeed, multigrid is known as one of the most cluster-friendly algorithms. Moreover, meta-algorithms such as multiple shooting \cite{bock1984multiple} try to break the spacetime optimization into a series of sub-optimizations that consider only a short animation segment and are thus faster to solve. Finally, we can also combine the benefits of both optimal and PD controllers by borrowing the idea of receding horizon control \cite{mayne1990receding}. In these controllers, optimal control is applied only to a short window of timesteps starting from the current one, and the window keeps being shifted forward to cover the whole animation.}

\appendix
\begin{changedBlk}
\section{KKT System of the \AbbrAO Subproblem}\label{appen:appenA}
We derive here the KKT system for the \AbbrAO subproblem. Instead of simply introducing the Lagrangian multipliers and following standard techniques as we did for the \AbbrPO subproblem, we present a derivation based on the analysis of the ghost force $u_i$. We first eliminate the Navier-Stokes constraints by writing $u_i$ as a function of $v_i$ and $v_{i+1}$. Next, we plug this function into our objective to obtain:
\begin{eqnarray*}
\frac{r}{2}\sum_{i=0}^{N-1}\|u_i(v_i,v_{i+1})\|^2+\frac{K}{2}\sum_{i=0}^{N-1}\|v_i-v_i^*\|^2.
\end{eqnarray*}
Taking the derivative of this objective against $v_i$ and considering the additional solenoidal constraints on $v_i$, we get the first two equations in $f$:
\begin{eqnarray*}
&&\frac{K}{r}(v_i-v_i^*)+\FPP{u_i}{v_i}^Tu_i+\FPP{u_{i-1}}{v_i}^Tu_{i-1}+\nabla\bar{p}_i=0    \\
&&\nabla\cdot v_i=0,
\end{eqnarray*}
where $\bar{p}_i$ is the Lagrangian multiplier. Now in order to derive the other two conditions in \prettyref{eq:KKTAO}, we need to determine the additional pressure $p_i$. We assert that $p_{i+1}$ is the Lagrangian multiplier of the solenoidal constraints on $u_i$. In fact, if $u_i$ is not divergence-free, we can always perform a pressure projection on $u_i$ by minimizing $\|u_i-\nabla p_{i+1}\|^2$ to get a smaller objective function value. As a result, $u_i$ must be divergence-free at the optima with $p_{i+1}$ being the Lagrangian multiplier, and we get the two additional equations of $f$:
\begin{eqnarray*}
&&\frac{v_{i+1}-v_i}{\Delta t}+\ASelf{v_{i+1}}-u_i+\nabla p_{i+1}=0 \\
&&\nabla\cdot u_i=0.
\end{eqnarray*}
From these two conditions, we can see that $\FPP{u_i}{v_i}=-\E{Q}\frac{I}{\Delta t}$, $\FPP{u_{i-1}}{v_i}=\E{Q}(\frac{I}{\Delta t}+\FPP{\ASelf{v_i}}{v_i})$. Here $\E{Q}$ is the solenoidal projection operator introduced in \prettyref{eq:KKTPO}. However, we can drop this $\E{Q}$ because we have $\FPP{u_{i-1}}{v_i}^Tu_i=(\frac{I}{\Delta t}+\FPP{\ASelf{v_i}}{v_i})^T\E{Q}^Tu_i$ and $\E{Q}^Tu_i=\E{Q}u_i=u_i$ by the fact that $u_i$ is already solenoidal.
\end{changedBlk}

\bibliographystyle{main}
\bibliography{main}
\end{document}